\documentclass[journal]{IEEEtran}

\ifCLASSINFOpdf
 
\else
 
\fi

\usepackage{graphicx}
\usepackage{amssymb}
\usepackage{amsmath}
\usepackage{ntheorem}
\usepackage{verbatim}
\usepackage{bbm}
\usepackage{amsbsy}
\newtheorem{myTheo}{Theorem}

\newtheorem{Lemma}{Lemma}
\newtheorem{Assumption}{Assumption}
\newtheorem{Remark}{Remark}

\newtheorem{Problem}{Problem}
\usepackage{floatrow}
\usepackage{graphicx}
\usepackage[label font=bf,labelformat=simple]{subfig}% <-- changed
\allowdisplaybreaks[4]

\begin{document}

\title{Non-singular Cooperative Guiding Vector Field Under a Homotopy Equivalence Transformation}

\author{Zirui Chen and Zongyu Zuo,~\IEEEmembership{Senior Member,~IEEE}% <-this % stops a space
\thanks{This work was supported by the National Natural Science Foundation of China under Grant 62073019.}
\thanks{Z. Chen and Z. Zuo are with Seventh Research Division, Beihang University (BUAA), Beijing,
 100191, China.  e-mail: chenzirui@buaa.edu.cn; zzybobby@buaa.edu.cn.}% <-this % stops a space
}

\markboth{IEEE TRANSACTIONS ON AUTOMATIC CONTROL}%
{Shell \MakeLowercase{\textit{et al.}}: Bare Demo of IEEEtran.cls for IEEE Journals}

\maketitle

\begin{abstract}
The present article advances the concept of a non-singular cooperative guiding vector field under a homotopy equivalence transformation. Firstly, the derivation of a guiding vector field, based on a non-singular vector field, is elaborated to navigate a transformed path from another frame. The existence of such vector fields is also deliberated herein. Subsequently, a coordination vector field derived from the guiding vector field is presented, incorporating an in-depth analysis concerning the impact of the vector field parameters. Lastly, the practical implementation of this novel vector field is demonstrated by its applications to 2-D and 3-D cooperative moving path following issues, establishing its efficacy.
\end{abstract}

\begin{IEEEkeywords}
path following, vector field, homotopy equivalence transformation, singularity, cooperation.
\end{IEEEkeywords}

\IEEEpeerreviewmaketitle

\section{Introduction}

\IEEEPARstart{A}{utonomous} robotic vehicles often perform trajectory tracking and path following tasks as part of their motion control obligations. In trajectory tracking, the vehicle has to adhere to a designated trajectory within time constraints. In path following, the vehicle must follow a predetermined path at a given speed, which may or may not involve time constraints \cite{Oliveira_2016}. In recent decades, the path following control idea is attracting growing attentions for its advantage in dealing with military and rescue tasks, including the monitoring of borders, reconnaissance of specified geographic areas, and safeguarding of convoys \cite{Wang_2019}.

In order to solve path following problem, several solution strategies have been proposed so far \cite{Sujit_2014}, such as ${H} _{2} /{H}_{\infty}$ controller design \cite{Ostertag_2008An}, integral line-of-sight control \cite{Caharija_2015A,Borhaug_2008Integral,Caharija_2016Integral}, nonlinear feedback control \cite{Lapierre_2007Nonlinear}, nonlinear model predictive control \cite{Faulwasser_2016Nonlinear}, guiding vector field control \cite{Nakai_2013Vector,Rezende_2022}. Besides, there is also some work in coordinated path following control \cite{Lan_2011Synthesis,Xie_2022Cooperative,Alessandretti_2020An} and formation path following control \cite{Eek_2021Formation}.

Among all the methods for studying the path following problem, the vector-field-based method has garnered extensive attention \cite{Yao_2021Singularity}. This approach has been successfully applied to various types of paths, including lines or circles \cite{Nelson_2007}, time-varying curves in $n$-Dimensions \cite{Goncalves_2010Vector}, bi-circular paths \cite{Liang_2014}, moving paths \cite{Kapitanyuk_2017}, and manifolds on Riemann space \cite{Yao_2023}. The vector field methodology is a global technique for solving the path following problem, avoiding singularity issues that arise from the conventional approach's use of the Frenet frame. However, despite its utility, there are inherent limitations to the vector field approach. Chief among these issues is the occurrence of singularity as a result of the zero vector \cite{Yao_2021Singularity}. Indeed, it has been demonstrated that if the vector field is to guide a path that is homologous to a circle or intersects itself, it must incorporate at least one singular point, with the measure of the singular set being zero \cite{Yao_2023}. While the analysis in \cite{Goncalves_2010Vector} assumes that the singular points are repulsive, this assumption is not upheld in \cite{Kapitanyuk_2018} for a planar desired path. Another challenge associated with the vector field approach is the lack of a parameter for cooperation, presenting difficulties in addressing cooperative path following problems.

Introducing an additional parameter has been identified as a potential solution for addressing the aforementioned challenges \cite{Yao_2021Singularity,yao2022guiding}. By altering the topological composition of the original path, it is possible to ensure that the guiding vector remains non-zero globally. Further, a coordinate proposal can be developed based on this supplementary parameter, ultimately enabling the construction of a distributed cooperative vector field.

In the absence of a guiding vector field, numerous consensus path-following algorithms have been suggested in the literature for multi-robot navigation. In an early attempt, \cite{Zuo_2015Three} presented an initial endeavor to provide a formal definition for the consensus path-following control problem in multi-agent systems. The main objective of this control problem is to guide a group of agents along a specific spatial path without any temporal constraints. To attain cooperative path following control for mobile robots, a framework has been proposed comprising of two distinctive loops. The outer loop addresses robot guidance, while the inner loop pertains to dynamics control, as stated in the literature \cite{Zuo_2022}. In this framework, the outer loop manages the angular speed, whereas the inner loop regulates the velocity of the non-holonomic wheeled mobile vehicle. A similar structure has been utilized in other studies \cite{Wang_2019}. Furthermore, the approach introduced in \cite{Oliveira_2016} applies a parameter updating rule and an angle control law to address the moving path following problem. A speed regulation law has also been designed to efficiently carry out the cooperative task.

However, the utilization of the Frenet frame method inevitably encounters singularities due to its dependence on the local structure of a single point on the path. The issue lies in the fact that the path following problem requires the robot to move along a geometric curve, rather than solely focusing on a single point on the path. Although the vector field approach effectively addresses the problem of local singularities, complications arise when the path undergoes time-varying transformations, as is the case with moving path following. The transformations between frames further make vector field calculations cumbersome. Thus, this paper aims to address these issues by focusing on constructing a non-singular cooperative guiding vector field under homotopy equivalence transformations. Furthermore, we aim to discuss the identity and features of this novel type of vector field.

Compared with some early attempts in \cite{Kapitanyuk_2017,Frew_2008Coordinated}, the focal contributions of this article from a technical standpoint are enumerated as follows:
\begin{enumerate}
    \item A non-singular and controllable vector field can be obtained through the application of a homotopy equivalence transformation. The existence of this vector field is contingent upon the invertibility of the Jacobian matrix associated with the transformation. The resulting vector field is capable of fulfilling navigation tasks even when a time-varying path is associated with a time-varying transformation rather than being dependent on time directly.
    \item The derived non-singular controllable vector field under transformation has been applied to the problem of cooperative moving path following, avoiding singularity issues caused by introducing the Frenet frame in traditional methods.
    \item The cooperative structure present in the cooperative transformation vector field is identified, enabling coordinated moving path tracking in multi-agent systems without requiring each individual to have direct knowledge of the specific details of the transformation. Furthermore, we engage in a discussion regarding the impact of the vector field parameters, concluding that the convergence parameter ought $g$ to be less or equal than $1$ while larger than $0$ when the coordination vector field is taken into account.
    \item The effectiveness of the proposed vector field is validated through the coordinated mobile path tracking problem.
\end{enumerate}

The subsequent sections of this paper are structured as follows. In Section \ref{section 2}, the fundamentals of graph theory and guiding vector fields for path following are introduced. Section \ref{section3} expounds on the extension of the guiding vector field for transformed situation and navigation along desired paths. Furthermore, in Section \ref{section4}, a cooperative guiding vector field under revertible transformation is proposed. Additionally, an application for cooperative moving path following (CMPF) problem is discussed in Section V. Finally, concluding remarks are presented in Section \ref{section 6}.

\section{Preliminaries}
\label{section 2}
\subsection{Notation}
In this paper, if the coordination $X$ is relative to frame $\{ I\}$, it would be symbolized by the left superscript $^IX$. The matrix or tensor $^PR_I$ represents it is a transformation from frame $\{ I\}$ to frame $\{ P\}$. For a system with $N$ robots, the superscript $(\cdot)^{[i]}$ represents any quantity that is connected to the robot indexed as $i$. The symbol ``$:=$" denotes the meaning "defined as". And If space $\mathcal{A}$ is homologous to $\mathcal{B}$, then $\mathcal{A} \approx \mathcal{B}$ holds. $\oplus$ means the direct sum of two transformation.
\subsection{Graph Theory}
This section adheres to the convention as specified in reference \cite{Olfati-Saber_2007}. A graph with $n$ items is denoted by $\mathcal{G}=(\mathcal{V},\mathcal{E})$. In this context, the set of vertices, denoted by $\mathcal{V} := \{1,\ldots,N\}$, is finite and contains a specific set of elements. The set of edges, denoted by $\mathcal{E}$, is a subset of $\mathcal{V}\times \mathcal{V}$ and consists of elements in the form of $(i, j)$, representing the adjacency between vertex $i$ and $j$. It is important to note that both $i$ and $j$ belong to $\mathcal{V}$. If the graph $\mathcal{G}$ is undirected, it is \emph{connected} if there exists a pathway between all vertices within the set $\mathcal{V}$. And for directed graph $\mathcal{G}$, it has a \emph{spanning tree} when there exists a pathway from the root to every vertices in the set $\mathcal{V}$. The adjacency matrix $A(\mathcal{G})$ of an undirected graph $\mathcal{G}$ is a symmetric $N \times N$ matrix used to encode the adjacency relationships between vertices. Its elements are defined as $[A(\mathcal{G})]_{ij} = 1$ if the vertices $(i, j)$ are connected by an edge in $\mathcal{G}$, and $[A(\mathcal{G})]_{ij} = 0$ otherwise. On the other hand, the Laplacian matrix $L(\mathcal{G})$ of $\mathcal{G}$ is an $N \times N$ matrix with $[L(\mathcal{G})]_{ij} = -a_{ij}$ if $i\neq j$ and $[L(\mathcal{G})]_{ii} = \sum_{k=1}^Na_{ik}$ for $i \leq n$. Here, $a_{ij}$ represents the $ij$th entry of the adjacency matrix. The matrix $D \in \mathbb{R}^{N\times \mathcal{E}}$  is an incidence matrix. In the case of an undirected graph, the determination of $D$ can be achieved through the allocation of arbitrary orientations to its edges.
\subsection{General Time-varying Guiding Vector Field}
In this section, we introduce the construction of a general time-varying Lyapunov guiding vector field. In \cite{Goncalves_2010Vector}, the desired path $\mathcal{P}$ is described as a set $\mathcal{T}(t)=\{X\in \mathbbm{R}^n|\alpha_i(X,t)=0,i \leq n-1\}$ of points lying in the intersection of level sets $\alpha_i = 0$ where $\alpha_i(X,t): \mathbbm{R}^{n+1}\mapsto \mathbbm{R}$.

Choose a positive definite Lyapunov function $V(\alpha_1,\alpha_2,...\alpha_{n-1},t)$, our aim is to design a vector field whose invariant set is $\mathcal{T}(t)$ and $\chi(t)$ remains on $\mathcal{T}(t)$ once $\chi(t_0) \in \mathcal{T}(t)$ for certain $t_0\geq 0$. Noticing the fact that the null space of a 1-D curve is spanned by a non-zero vector $\mathbf{n}_0$. When the vector field goes parallel with $\mathbf{n}_0$, $\alpha_i$ remains a constant value while the vector is non-zero. This proposition helps to determine the moving direction of the vector field.

The time derivative of $V(\alpha_1,\alpha_2,...\alpha_{n-1},t)$ is
\begin{equation}
    \label{2c-vdot1}
    \dot{V} = \nabla V^{\top} \dot{X} + \frac{\partial V}{\partial t}
\end{equation}
The work \cite{Goncalves_2010Vector} gives a general expression of vector field for $\dot{X}=u$ satisfying the navigation goal
\begin{equation}
    \label{goncalves vector field}
    u = -G\nabla V +H (\wedge_{i=1}^{n-1}\nabla \alpha_i) - (\nabla V^{\top})^{+}\frac{\partial V}{\partial t}
\end{equation}
where $G$ and $H$ are diagonal positive definite matrices, $(\cdot)^+$ denotes the generalized inverse of a non-square matrix. (\ref{goncalves vector field}) derives $\dot{V} = -G \Vert V \Vert^2$, Indicating that $V$ would becomes $0$. So the position of the robot would finally satisfy $\alpha_i = 0$ for $i = 1,\ldots,n-1$, meaning the trajectory of the robot would converge to $\mathcal{T}(t)$.

Moreover, when the desired path $\mathcal{P}$ is time-invariant, and the Lyapunov function is chosen as $V(\alpha_1,\alpha_2,...\alpha_{n-1}) = \sum_{i=1}^{n-1}k_i\frac{1}{2}\alpha_{i}^2$, where $k_i$ is a positive constant for $k=1,\ldots,n-1$, the vector field (\ref{goncalves vector field}) could be simplified as
\begin{equation}
    \label{yao vector field}
    u = \wedge_{i=1}^{n-1}(\nabla\alpha_1,...,\nabla\alpha_{n-1})-\sum_{i=1}^{n-1}k_i \alpha_i \nabla \alpha_i
\end{equation}
And (\ref{yao vector field}) is used in \cite{Yao_2021Singularity,Yao_2023,yao2022guiding}.
\begin{Lemma}
    \label{wedge orthogonality}
    $\wedge_{i=1}^{n-1}(p_1,...,p_{n-1})$ exhibits orthogonality with respect to each of the vectors $p_1, \ldots, p_{n-1}$.
\end{Lemma}
\section{Non-singular Guiding Vector Field under Transformation for Single Robot}
\label{section3}
Singularity is a prevalent issue encountered in guiding vector field approaches. A straightforward explanation for this phenomenon is that at a singular point, there exist multiple tendencies to follow a particular direction to reach the desired path. For instance, when the desired path is a circle, the singular point corresponds to its center point. At this location, the distances between any point on the circle and the center point are identical, resulting in an equal tendency to reach every point on the circle. Therefore, the guiding vector becomes a zero vector to represent this equality. A similar situation arises at the intersection point of a self-intersecting curve. It has been demonstrated that this singularity always exists for a bounded curve homologous to a circle \cite{Yao_2023}. However, this singularity does not occur when the path is homologous to a straight line.

Based on this observation, a likely approach to emit the singularity is to "stretch" the bounded curve into an unbounded line. Intuitively, this particular operation could be viewed as disrupting the symmetrical properties of the original path, thus providing a reference direction for the singular point.
%\begin{figure}[htb]
%    \centering
%    \includegraphics[width=0.45 \linewidth]{f3-1-1.jpg}
%    \includegraphics[width=0.45 \linewidth]{f3-1-2.jpg}
%    \caption{An planar curve homologous to a circle could be homologous to a line in 3-D after "stretching".}
%    \label{fig:3-1-1}
%\end{figure}

\begin{figure*} [!ht]
	\centering
	\subfloat[\label{fig:1a}]{
		\includegraphics[width=0.5 \linewidth]{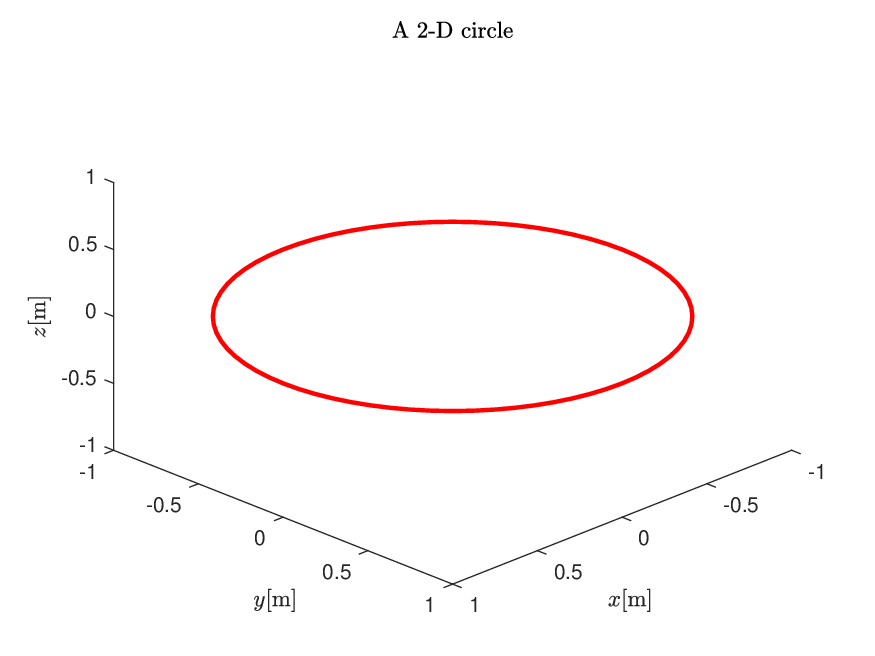}}
	\subfloat[\label{fig:1b}]{
		\includegraphics[width=0.5 \linewidth]{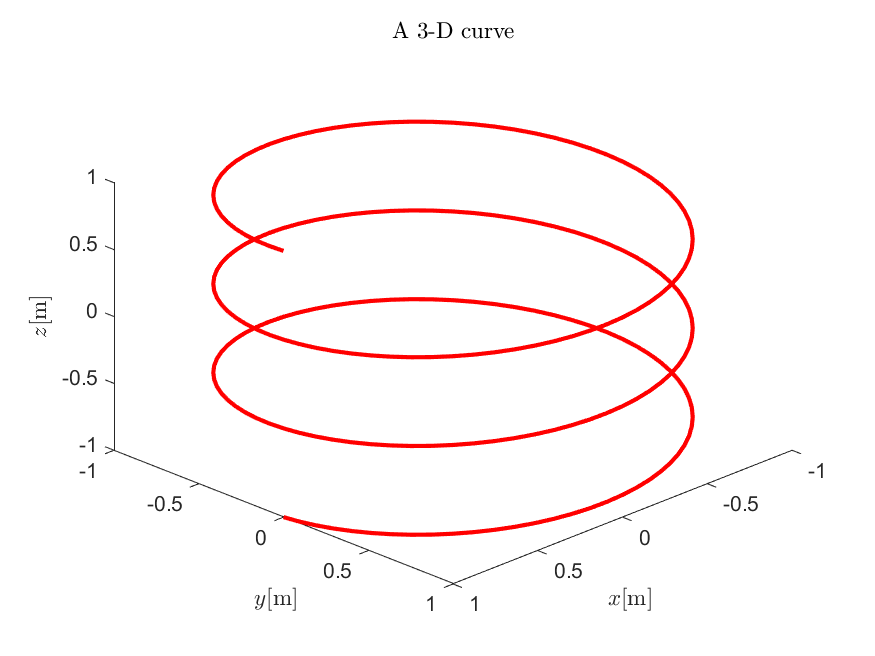}}
	\\
	\subfloat[\label{fig:1c}]{
		\includegraphics[width=0.5 \linewidth]{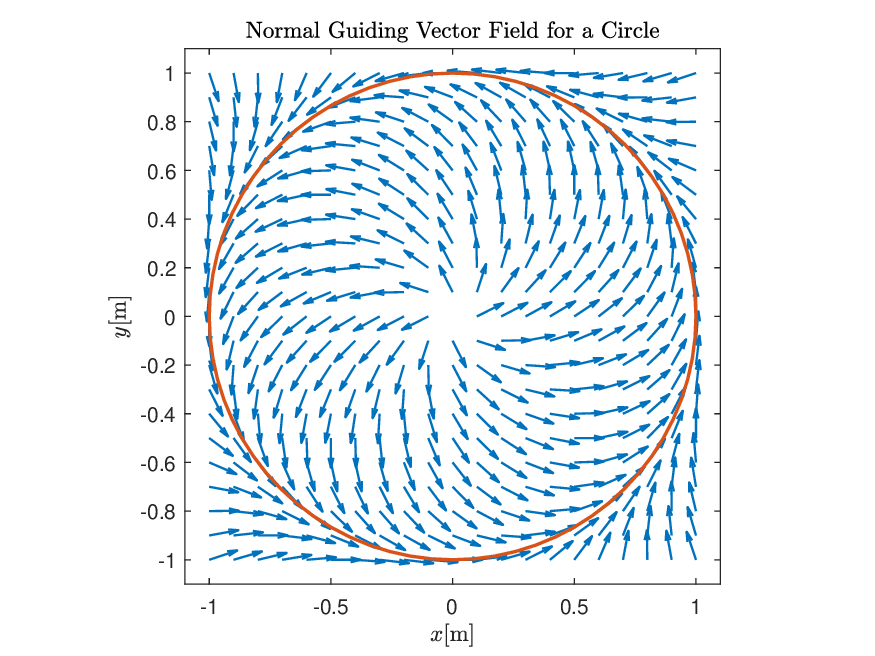} }
	\subfloat[\label{fig:1d}]{
		\includegraphics[width=0.5 \linewidth]{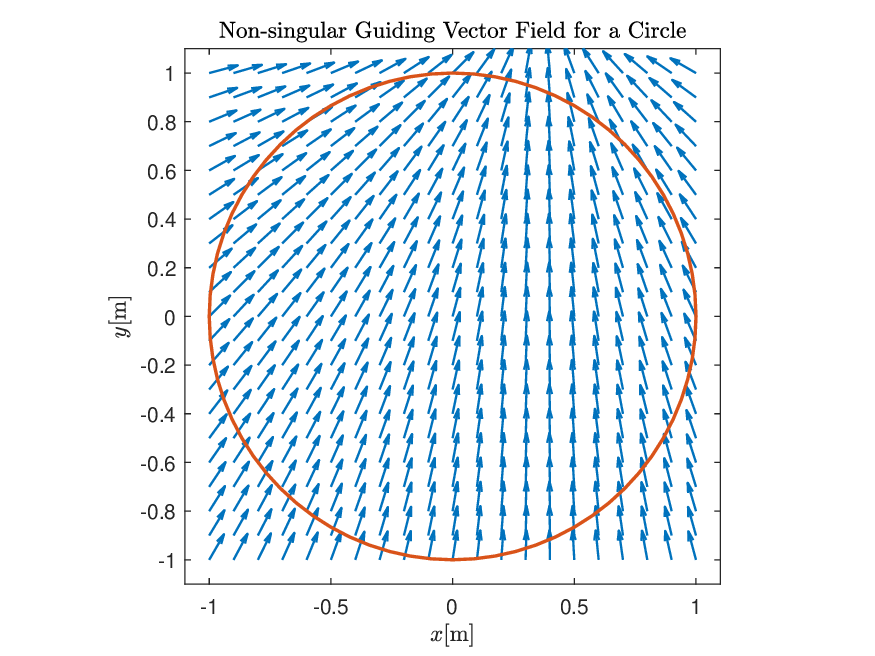}}
	\caption{An planar curve homologous to a circle (\ref{fig:1a}) could be homologous to a line in 3-D (\ref{fig:1b}) after "stretching". (\ref{fig:1c}): the corresponding normal guiding vector field for (\ref{fig:1a}). (\ref{fig:1d}): the corresponding non-singular guiding vector field for (\ref{fig:1b}) }
	\label{fig31}
\end{figure*}

\begin{Remark}
    The primary reason why the curve can change its topological identity is that the one-dimensional (1-D) manifold can be considered as a single-parameter transformation group. This implies that there exists a point $\theta(t) \in \mathcal{S}$ moving along the path, and the path can be regarded as the collection of the trajectory produced by the motion of $\theta(t)$. The extended curve can be expressed as $\mathcal{P}\times \mathcal{S}$. In the case of the (a) manifold in Fig. \ref{fig31}, $\mathcal{S}$ is homologous to a curve, so $\theta(t)$ returns to the same point after traversing $\mathcal{S}$. In contrast, for the (b) manifold, $\mathcal{S}$ is homologous to a straight line, which means $\theta(t)$ moves ahead and never returns. From this perspective, the original path may be perceived as a projection from $\mathcal{P}\times\mathcal{S}$ to $\mathcal{P}$.
\end{Remark}

It is imperative to elucidate the correlation between the initial path and state and the extended path and state.

For original path $\mathcal{P} \in \mathbbm{P}$, the extended path $\mathcal{P}^{ex} := \mathcal{P}\times\mathcal{S}$. So the extended path is the product of the original path and the set where the extra parameter $\theta$ in. For a point $^IX \in \mathbbm{I}$, the corresponding extended state is $\xi := (^IX^{\top}, \theta)^{\top}$.

Reversely, the original path $\mathcal{P}$ can be regarded as a projection from $\mathcal{P}^{ex}$ to it, mathematically
\begin{equation}
    \label{Pi}
    \Pi: \mathbbm{P} \times \mathcal{S} \rightarrow \mathbbm{P}, \  \Pi(\mathcal{P}^{ex}) = \mathcal{P}
\end{equation}
Also, the original state $^IX$ of the robot has a similar relationship with the extended states
\begin{equation}
    \label{pi}
    \pi: \mathbbm{P} \times \mathcal{S} \rightarrow \mathbbm{P} , \ \pi \left(\xi(^IX^{\top}(t),\theta(t))^{\top} \right) = ^IX
\end{equation}
For $^IX(t)$ and $\theta(t)$ are function depending on $t$, $\xi$ also depends on $t$, denoting by $\xi(t)$.

Since the fundamental concept of a non-singular vector field has already been introduced, the primary concern of this section is to address the main issue at hand, which is:
\begin{Problem}
    \label{problem1}
    For a homotopy equivalence transformation $\mathcal{F}: \mathbbm{I} \mapsto \mathbbm{P} $ and a time-invariant path $\mathcal{P}$ in $\mathbbm{P}$, find a vector field $\chi: \mathbbm{I}\times \mathcal{S} \mapsto T_P(\mathbbm{I}\times \mathcal{S})$ for the equation $\dot{\xi} = \chi(\xi(t))$, such that the following two conditions are satisfied
    \begin{enumerate}
        \item  $\Lambda_+$ is the invariant set of the system $\dot{\xi} = \chi(\xi(t))$, $\mathcal{F}(\Pi(\Lambda_+)) = \mathcal{P}$,
        \item For every $^IX \in \mathcal{F}^{-1}(\mathcal{P})$, $\chi(\xi(^IX,\theta))$ is non-zero.
    \end{enumerate}
\end{Problem}

\begin{Lemma}
    \label{extended homotopy equivalence transformation}
    Let $\mathcal{F}:\mathbbm{I} \rightarrow \mathbbm{P}$ be a homotopy equivalence transformation and $\mathcal{A}:\mathbbm{S} \rightarrow \mathbbm{S}$ be the identical transformation. Suppose that the spaces operated by $\mathcal{F}$ and $\mathcal{A}$ are disjoint. Then, the direct sum of $\mathcal{F}$ and $\mathcal{A}$, denoted as $\mathcal{F}^{ex} = \mathcal{F} \oplus \mathcal{A}: \mathbbm{I}\times\mathbbm{S}\rightarrow \mathbbm{P}\times\mathbbm{S}$, is also a homotopy equivalence transformation.
\end{Lemma}

\begin{IEEEproof}
    \label{proof for extended homotopy equivalence transformation}
    See appendix A.
\end{IEEEproof}

\begin{figure}[H]
    \centering
    \includegraphics[width=1 \linewidth]{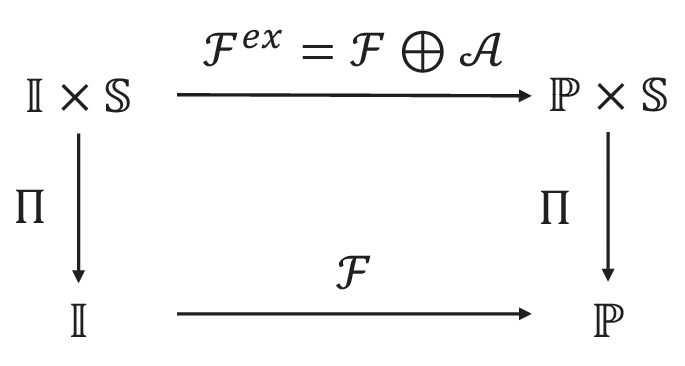}
    \caption{Relationships between the spaces and transformations in this paper}
    \label{fig:3-2}
\end{figure}

In order to enhance clarity, a visual representation of the relationships between the spaces and transformations discussed in this paper is presented in Fig. \ref{fig:3-2}.

In the present study, it is imperative to assert that the transformation under investigation exhibits temporal variation aligning with $\zeta(t)$, $^PX = \mathcal{F}(^IX,\zeta(t))$, mathematically. Consequently, the path in question may be considered relatively static with respect to a moving frame. In studies such as \cite{Kapitanyuk_2017}, it is assumed that the kinetic differential equation is integrable, leading to an explicit dependence of the path expression on time $t$. However, in cases where the moving frame undergoes motion with non-holonomic constraints, the trajectory becomes non-integrable, making it impossible to obtain such an expression. As a result, the vector field (\ref{goncalves vector field}) cannot be directly constructed. Moreover, the invariant set of normal vector field is the union of the desired path and singular points, meaning the guiding vector field is ineffective on these singular points.

As in \cite{Yao_2021Singularity}, the desired path  $\mathcal{P} \in \mathbbm{P}$ is described by parameter $\theta \in \mathcal{S}$
\begin{equation}
    \label{path parameter equation}
    ^Px_1 = f_1(\theta), \  ^Px_2 = f_2(\theta), ..., \  ^Px_n = f_n(\theta)
\end{equation}
\begin{Assumption}
    \label{bounded derivative of path}
    The first and second derivatives of $f_i$ are bounded for every $i = 1,\ldots,n$.
\end{Assumption}
So the extended manifold $^h\mathcal{P}$ in $\mathbbm{P}\times \mathcal{S}$ could be regarded as the intersection of the following hyperplane
\begin{equation}
    \label{path extended equation}
    \phi_1 = {}^Px_1 - f_1(\theta),...,  \  \phi_n = {}^Px_n - f_n(\theta)
\end{equation}
\begin{Remark}
    For convenience, in this paper, some notations are defined as follows: \\
    $\nabla^I V:= \left[ \begin{array}{ccc}  \frac{\partial V}{\partial ^Ix_1} & \ldots & \frac{\partial V}{\partial ^Ix_n} \end{array}\right]^{\top}$, \\
    $\nabla^P V:= \left[ \begin{array}{ccc}  \frac{\partial V}{\partial ^Px_1} & \ldots & \frac{\partial V}{\partial ^Px_n} \end{array}\right]^{\top}$, \\
    $\nabla^{\xi} V:= \left[ \begin{array}{cccc}  \frac{\partial V}{\partial ^Px_1} & \ldots & \frac{\partial V}{\partial ^Px_n} & \frac{\partial V}{\partial \theta} \end{array}\right]^{\top}$,\\
    $\Phi := (\phi_1,\phi_2,\ldots,\phi_n)^{\top}$, \\
    $^PX:=({}^Px_1,{}^Px_2,\ldots,{}^Px_n)^{\top}$.
\end{Remark}

The candidate positive definite Lyapunov function $V = V(\Phi)$, and the time derivative of $V$ is

    \begin{align}\label{Vdot30}
        \dot{V} &=\sum_{i=1}^{n}\frac{\partial V}{\partial \phi_i} \dot{\phi}_i  \nonumber\\
                &=\left(\frac{\partial V}{\partial \Phi}\right)^{\top} \begin{bmatrix}
                    \nabla^{\xi}\phi_1^{\top} \\
                    \nabla^{\xi}\phi_2^{\top} \\
                    \vdots \\
                    \nabla^{\xi}\phi_n^{\top} \\
                    \end{bmatrix}\begin{bmatrix}
                    ^P\dot{x}_1 \\
                    ^P\dot{x}_2 \\
                    \vdots \\
                    ^P\dot{x}_n \\
                    \dot{\theta}
    \end{bmatrix}
    \end{align}
One obtains from (\ref{path extended equation}) that $\nabla^{\xi} \phi_i=\left(0, \ldots, 1, \ldots,-\frac{\partial f_i}{\partial \theta}\right)^{\top}$ for $i = 1,\ldots,n$, and $1$ is the $i$th element of the gradient vector. So the compact form of (\ref{Vdot30}) is derived as

 \begin{align}   \label{Vdot31}
            \dot{V} &=(\frac{\partial V}{\partial \Phi})^{\top}\left[\begin{array}{ccccc}
            1 & 0 & \cdots & 0 & -\frac{\partial f_1}{\partial \theta} \\
            0 & 1 & \cdots & 0 & -\frac{\partial f_2}{\partial \theta} \\
            \vdots & \vdots & \ddots & 0  & \vdots \\
            0 & 0 & \cdots & 1 & -\frac{\partial f_n}{\partial \theta}
        \end{array}\right]\left[\begin{array}{c}
             ^P\dot{x}_1 \\
             ^P\dot{x}_2 \\
             \vdots \\
             ^P\dot{x}_n \\
             \dot{\theta}
        \end{array} \right]\nonumber \\
        &=(\frac{\partial V}{\partial \Phi})^{\top} \left(\left[\begin{array}{c}
             ^P\dot{x}_1 \\
             ^P\dot{x}_2 \\
             \vdots \\
             ^P\dot{x}_n
        \end{array} \right] - \left[\begin{array}{c}
            \frac{\partial f_1}{\partial \theta} \\
            \frac{\partial f_2}{\partial \theta} \\
            \vdots \\
            \frac{\partial f_n}{\partial \theta}
        \end{array}\right] \dot{\theta}\right)
\end{align}
It is noteworthy to highlight that $^PX$ part and $\theta$ part are separately in (\ref{Vdot31}), signifying that $\theta$ is an extensive parameter.  \\

While the transformation from $\mathbbm{I}$ to $\mathbbm{P}$ is
\begin{equation}
    \label{F}
    ^PX = \mathcal{F}({}^IX,\zeta(t))
\end{equation}
whose time derivative is
\begin{equation}
    \label{Fdot}
    ^P\dot{X} = \nabla\mathcal{F}^I\dot{X}+\frac{\partial \mathcal{F}}{\partial \zeta}\dot{\zeta}(t)
\end{equation}
where $\nabla\mathcal{F} = \frac{\partial ^PX}{\partial ^IX}$ is the \emph{Jacobian} between $\mathbbm{I}$ and $\mathbbm{P}$, usually denoted by $\mathbf{J}$ in differential geometry. According to (\ref{Fdot}), if $\nabla \mathcal{F}$ is invertible, a fully time compensation could be accomplished by
\begin{equation}
    \label{chix}
    \chi_0 = {}^I\dot{X} = (\nabla \mathcal{F})^{-1} \left( \pmb{u} - \frac{\partial \mathcal{F}}{\partial \zeta} \dot{\zeta}(t)\right)
\end{equation}
where $\pmb{u} \in \mathbbm{R}^n$.
\begin{Remark}
    The transformation studied in this paper is influenced by time $t$ through an indirect correlation with $\zeta(t).$ In practical scenarios, \emph{section \ref{section5}} for example, $\zeta(t)$ may refer to the position of the leader target or another physical property. (\ref{chix}) indicates that this information can be obtained through an internal model, allowing the controlled agent to successfully achieve the path following task without direct knowledge of the leader's exact information. This issue would be discussed deeper in \emph{section \ref{51b}}.
\end{Remark}

Combining (\ref{Vdot31}), (\ref{F}), (\ref{Fdot}) and (\ref{chix}), the time derivative of $V$ can be simplified as
\begin{align} \label{vdot32}
    \dot{V} &= (\frac{\partial V}{\partial \Phi})^{\top} \left(\pmb{u} + \left[\begin{array}{c}
        \frac{\partial f_1}{\partial \theta} \\
        \frac{\partial f_2}{\partial \theta} \\
        \vdots \\
        \frac{\partial f_n}{\partial \theta}
    \end{array}\right] \dot{\theta}\right) \nonumber\\
    & = (\frac{\partial V}{\partial \Phi})^{\top} \left[\begin{array}{c}
         \nabla^{\xi}\phi_1^{\top} \\
         \nabla^{\xi}\phi_2^{\top} \\
         \vdots \\
         \nabla^{\xi}\phi_n^{\top} \\
    \end{array} \right]\left[\begin{array}{c}
         u_1 \\
         u_2 \\
         \vdots \\
         u_n \\
         \dot{\theta}
    \end{array} \right]
\end{align}
So the term varying over time due to the time-varying transformation has been fully emitted.
\begin{Remark}
    From (\ref{chix}), it can be seen that a prerequisite for compensation to exist is the invertibility of $\nabla \mathcal{F}$. This observation highlights the possibility of constructing this type of vector field not only under rotational or translational transformations (both of them are special cases of orthogonal congruent transformations) but also under any invertible transformation.
\end{Remark}

Based on (\ref{vdot32}), we can construct a guiding vector field (\ref{goncalves vector field}) or (\ref{yao vector field}) to accomplish the path following task. An example is given in \emph{Section \ref{section5}}. The designed vector field solving Problem \ref{problem1} is:
\begin{align}\label{chimpf}
        \chi_{\mathrm{mpf}} = &\left[\begin{array}{cc}
      (\nabla\mathcal{F})^{-1}   &  \mathbf{0}_{n\times1} \\
       \mathbf{0}_{1\times n}  & 1
    \end{array} \right] \left( -G\nabla^P V + H \wedge_{i=1}^{n-1}\nabla^P\phi_i\right)  \nonumber\\
    & -\left[\begin{array}{c}
         (\nabla\mathcal{F})^{-1} \frac{\partial \mathcal{F}}{\partial \zeta} \dot{\zeta}(t) \\
         0
    \end{array}\right]
\end{align}

\begin{myTheo}
    A solution for Problem \ref{problem1} is (\ref{chimpf}) under \emph{Assumption \ref{assumption1}}.
\end{myTheo}
\begin{IEEEproof}
   Equation  (\ref{vdot32}) could be rewritten as
    \begin{equation}
        \label{vdot33}
        \dot{V} = (\nabla^{\xi} V)^{\top}  \left[\begin{array}{c}
         u_1 \\
         u_2 \\
         \vdots \\
         u_n \\
         \dot{\theta}
    \end{array} \right]
    \end{equation}
    Applying (\ref{chimpf}) to (\ref{Vdot31}), and combining with (\ref{vdot33}), one obtains
    \begin{equation}
        \label{vdot34}
        \dot{V} = -G \Vert \nabla^{\xi} V \Vert^2
    \end{equation}
    Together with the definition of $V$, According to \emph{Theorem 4.10} in \cite{khalil2002nonlinear}, the system $\dot{\xi} = \chi_{\mathrm{mpf}}(\xi(t))$ is global exponential stable. This implies every $\xi(t_0)$ would converge to the invariant set of this system. Since $\mathcal{F}$ is a homotopy equivalence transformation, so its extension $\mathcal{F}^{ex}$ is also a homotopy equivalence transformation according to \emph{Lemma} \ref{extended homotopy equivalence transformation}. Thus, the extended desired path in $\mathbbm{I}$ satisfies $(\mathcal{F}^{ex})^{-1}(\mathcal{P}^{ex}) \approx \mathbbm{R} $. \emph{Theorem 2} in \cite{Yao_2021Singularity} has proven that no singular point exists when the desired path is homologous to $\mathbbm{R}$, the singular set $\mathcal{C} = \varnothing$. According to \emph{Lemma 2} in \cite{Yao_2021Singularity}, the invariant set $\Lambda_+$ is $(\mathcal{F}^{ex})^{-1}(\mathcal{P}^{ex})$, $\Lambda_+ = (\mathcal{F}^{ex})^{-1}(\mathcal{P}^{ex})$ mathematically. Then one obtains $\mathcal{F}^{ex}(\Lambda_+) = \mathcal{P}^{ex}$. So $\Pi(\mathcal{F}^{ex}(\Lambda_+)) = \mathcal{P}$. Since the extension part is separated from the original part, and $\Pi$ emits it, one can derive $\mathcal{F}(\Pi (\Lambda_+))=\mathcal{P}$. Thus \emph{1)} in \emph{Problem \ref{problem1}} holds.

    From (\ref{path parameter equation}), one calculates that $\nabla^{\xi} \phi_i=\left(0, \ldots, 1, \ldots,-\frac{\partial f_i(\theta)}{\partial \theta}\right)^{\top}$ for $i=1, \ldots, n$, where 1 is the $i$-th component of the gradient vector. Therefore,
        $$
        \wedge_{i=1}^{n-1}\nabla^{\xi}\phi_i=(-1)^n\left[\begin{array}{c}
        \frac{\partial f_1(\theta)}{\partial \theta} \\
        \vdots \\
        \frac{\partial f_n(\theta)}{\partial \theta} \\
        1
        \end{array}\right] \in \mathbb{R}^{n+1}
        $$
    It's clear that $\wedge_{i=1}^{n-1}\nabla^{\xi}\phi_i$ is not equal to zero vector. From \emph{Lemma} \ref{wedge orthogonality}, one obtains that the linear combination of $\wedge_{i=1}^{n-1}\nabla^{\xi}\phi_i$ and $\nabla^{\xi}\phi_i$ for $i=1,\ldots,n-1$ equals to zero if and only if all these vectors are zero vector, conflicting to the fact that $\wedge_{i=1}^{n-1}\nabla^{\xi}\phi_i$ is unequal to zero vector. So for any $\xi$, $\chi(\xi) \neq 0$, the second requirement in \emph{Problem \ref{problem1}} is satisfied.
\end{IEEEproof}
\section{Distributed  Cooperative Guiding Vector Field}
\label{section4}
It is important to note that the virtual coordinates' coordination has a direct impact on the robotic motions' coordination, as the virtual coordinate corresponds to a certain parameter of the desired path, such as the natural parameter. Additionally, equation (\ref{Vdot31}) indicates that the supplementary parameter, $\theta, $ is disentangled from the actual coordination component. Consequently, the collaboration among robots may be achieved through a consensus protocol tailored for $\theta.$

In order to achieve cooperation among robots in a network, it is essential to establish a desired pattern. This entails creating a desired pattern, denoted as $\Delta^{[i,j]}$, derived from a specified reference configuration $\Theta^* :=(\theta^{[1]*},\ldots,\theta^{[N]*})^{\top}$. Specifically, the vector stack of $\Delta^{[i,j]}$ is represented by $\Delta^{*}$, which is obtained by transposing $D$ with $\Theta^*$.

Before proposing the consensus algorithm, it is necessary to formulate the cooperative guiding vector field for path following problem as follows.
\begin{Problem}
    \label{problem2}
    For a smooth reversible transformation $\mathcal{F}: \mathbbm{I} \mapsto \mathbbm{P} $ and a time-invariant path $\mathcal{P}$ in $\mathbbm{P}$. Devise a coordinated guiding vector field $\mathfrak{X}^{[i]}$ for $i = 1,\ldots,n-1$ to ensure that the trajectories of $\dot{\mathbf{\xi}}^{[i]} = \mathfrak{X}^{[i]}(\mathbf{\xi}^{[i]},\zeta(t))$, having an initial condition $ \mathbf{\xi}^{[i]} \in R^{n+1}$ at $t = t_0 \geq 0$, satisfy the subsequent two control objectives.
    \begin{enumerate}
        \item  $\Lambda_+$ is the invariant set of the system $\dot{\mathbf{\xi}}^{[i]} = \mathfrak{X}^{[i]}(\mathbf{\xi}^{[i]},\zeta(t))$, $\mathcal{F}(\Pi(\Lambda_+)) = \mathcal{P}$. and for every $^Ix^{[i]} \in \mathcal{F}^{-1}(\mathcal{P})$, $\mathfrak{X}^{[i]}$ is non-zero.
        \item The motion of each robot is coordinated in a distributed manner, based on the communication graph $\mathcal{G}$, whereby communication between Robot $i$ and Robot $j$ only occurs if the pair $(i, j) \in \mathcal{E}$. The coordination ensures that the virtual coordinates of the robots satisfy $\theta^{[i]}(t) - \theta^{[j]}(t) - \Delta[i,j] \rightarrow 0$ as time approaches infinity, for all pairs $(i, j) \in \mathcal{E}$.

    \end{enumerate}
\end{Problem}

The cooperative guiding vector field we design here has the formulation as
\begin{equation}
    \label{cr vector in s4}
    \chi_{\mathrm{cr}}^{[i]}(\Theta) = \left(0, \ldots, 0,c^{[i]}(\Theta)\right)^{\top}
\end{equation}
where $\Theta = (\theta^{[1]},\ldots,\theta^{[n]})^T$ and $c^{[i]}$ is the consensus protocol. The consensus objective here is $\lim_{t\rightarrow \infty} (\theta^{[i]} - \theta^{[j]}-\Delta[i,j]) = 0$ for every $(i,j) \in \mathcal{E}$. We would like to suggest utilizing the subsequent consensus control algorithm \cite{yao2022guiding}:
\begin{equation}
    \label{consensus proposal}
    c^{[i]}=-\sum_{j\in \mathcal{N}_i}(\theta^{[i]} - \theta^{[j]}-\Delta^{[i,j]}) \ \  \forall (i,j) \in \mathcal{E}
\end{equation}
Thus the compact form of the consensus protocol is $\mathbf{c}(\Tilde{\Theta}) = -L\Tilde{\Theta}$, where $L$ is the Laplacian matrix. And the cooperative guiding vector field is $\mathfrak{X}^{[i]}=\chi_{\mathrm{mpf}}^{[i]}+\chi_{\mathrm{cr}}^{[i]}(\Theta)$.
Here an assumption is employed to guarantee the consensus when applying protocol (\ref{consensus proposal}).

\begin{Assumption}
     \label{assumption1}
%   The communication graph $\mathcal{G} = (\mathcal{V},\mathcal{E})$ is directed and has a spanning tree.
The communication graph $\mathcal{G} = (\mathcal{V},\mathcal{E})$ is undirected and connected.
\end{Assumption}

Assumption \ref{assumption1} ensures the consensus protocol works, which has been proven in \cite{Olfati-Saber_2007}.

So far, we could propose the cooperative guiding vector field
\begin{equation}
    \label{mathfrakX}
    \mathfrak{X}^{[i]} = \chi_{\mathrm{mpf}}^{[i]} + \chi_{\mathrm{cr}}^{[i]}
\end{equation}
It has been discovered that the two terms function independently. And in particular, $\chi_{\mathrm{cr}}^{i}$ solely operates in relation to the supplementary parameter $\theta^{[i]}$.
\begin{myTheo}
  problem \ref{problem2} is solved by (\ref{chimpf}), (\ref{cr vector in s4}) and (\ref{mathfrakX}) under \emph{Assumption} \ref{bounded derivative of path} and \ref{assumption1} if the last item $g$ of $G$ in (\ref{chimpf}) satisfying $g \geq 1$.
\end{myTheo}
\begin{IEEEproof}
    The time derivative of the extra parameter $\theta^{[i]}$ is the last row of (\ref{mathfrakX}), which is
    \begin{equation}
        \label{derivate of theta}
            \dot{\theta}^{[i]} = g\frac{\partial V^{[i]}}{\partial \theta^{[i]}} +(-1)^n + k_c c^{[i]}(\Theta)
    \end{equation}
    where $g$ is the last item in $G$ and $H$ is set to be an identical matrix of $n\times n$ for convenience. So for $\Tilde{\Theta}=\Theta-\Theta^*$, it holds
    \begin{equation}
        \label{derivative of Theta}
            \dot{\Tilde{\Theta}} = (-1)^n \mathbf{1}_N + g\left[\begin{array}{c}
                 \frac{\partial V^{[1]}}{\partial \theta^{[1]}} \\
                 \vdots \\
                 \frac{\partial V^{[N]}}{\partial \theta^{[N]}}
            \end{array}\right] + k_c \mathbf{c}(\Theta)
    \end{equation}
    noting that $D^{\top}\mathbf{1}_N = 0$, equation (\ref{derivative of Theta}) becomes (\ref{sim derivative of Theta}) when multiplying $D^{\top}$ at left:
    \begin{equation}
        \label{sim derivative of Theta}
        D^{\top}\dot{\Tilde{\Theta}} = gD^{\top}\left[\begin{array}{c}
                 \frac{\partial V^{[1]}}{\partial \theta^{[1]}} \\
                 \vdots \\
                 \frac{\partial V^{[N]}}{\partial \theta^{[N]}}
            \end{array}\right] - k_c D^{\top}L\Tilde{\Theta}
    \end{equation}
    Set the candidate Lyapunov function for the whole system as
\begin{align}\label{V41}
            \mathbbm{V} &= \sum_{i=1}^{N} V^{[i]} + \frac{1}{2}k_c \Tilde{\Theta}^{\top}L\Tilde{\Theta}\nonumber \\
            &= \sum_{i=1}^{N} V^{[i]} + \frac{1}{2}k_c (D^{\top}\Tilde{\Theta})^{\top} (D^{\top}\Tilde{\Theta})
\end{align}
where the fact $L = DD^{\top}$ has been applied here. Taking the time derivative of $\mathbbm{V}$, one derives
\begin{align}\label{V4dot0}
            \dot{\mathbbm{V}}=& -\sum_{i=1}^N G \Vert \nabla^{\xi} V^{[i]} \Vert^2 + k_c\sum_{i=1}^N (\nabla^{\xi} V^{[i]})^{\top} \left[ \begin{array}{c}
                 0\\
                 0\\
                 \vdots \\
                 c^{[i]}(\theta^{[i]})
            \end{array} \right] \nonumber\\
            & + k_c \left(g D^{\top} \left[\begin{array}{c}
                 \frac{\partial V^{[1]}}{\partial \theta^{[i]}}\\
                 \vdots\\
                 \frac{\partial V^{[n]}}{\partial \theta^{[n]}}
            \end{array} \right] - k_c D^{\top}L\Tilde{\Theta}\right)^{\top} D^{\top} \Tilde{\Theta}\nonumber \\
            = & -\sum_{i=1}^N G \Vert \nabla^{\xi} V^{[i]} \Vert^2 +2gk_c  \left[\begin{array}{ccc} \frac{\partial V^{[1]}}{\partial \theta^{[i]}} & \ldots&\frac{\partial V^{[n]}}{\partial \theta^{[n]}}\end{array} \right] L \Tilde{\Theta} \nonumber\\
            & -k_c^2 \Vert L \Tilde{\Theta} \Vert^2 \nonumber\\
            = & -\sum_{i=1}^N G \Vert \nabla^P V^{[i]} \Vert^2  \nonumber\\
            &- g \left[\begin{array}{ccc} \frac{\partial V^{[1]}}{\partial \theta^{[i]}} & \ldots&\frac{\partial V^{[n]}}{\partial \theta^{[n]}}\end{array} \right] \left[\begin{array}{c}
                 \frac{\partial V^{[1]}}{\partial \theta^{[i]}}\\
                 \vdots\\
                 \frac{\partial V^{[n]}}{\partial \theta^{[n]}}
            \end{array} \right]\nonumber \\
            &+2gk_c \left[\begin{array}{ccc} \frac{\partial V^{[1]}}{\partial \theta^{[i]}} & \ldots&\frac{\partial V^{[n]}}{\partial \theta^{[n]}}\end{array} \right] L \Tilde{\Theta}  -k_c^2 \Vert L \Tilde{\Theta} \Vert^2
        \end{align}
let $\alpha := \left[\begin{array}{ccc} \frac{\partial V^{[1]}}{\partial \theta^{[i]}} & \ldots&\frac{\partial V^{[n]}}{\partial \theta^{[n]}}\end{array} \right]^{\top}$ and $\beta := k_c^2 L \Tilde{\Theta}$, then (\ref{V4dot0}) becomes
    \begin{equation}
        \label{V4dot1}
            \begin{aligned}
             \dot{\mathbbm{V}}=& -\sum_{i=1}^N G \Vert \nabla^{\xi} V^{[i]} \Vert^2 - g \Vert \alpha \Vert^2 +2g\alpha^{\top} \beta -\Vert \beta \Vert^2 \\
             =& -\sum_{i=1}^N G \Vert \nabla^{\xi} V^{[i]} \Vert^2 - \left[ \begin{array}{cc} \alpha^{\top}   &  \beta^{\top} \end{array}\right] \left[\begin{array}{cc}
                 g & -g \\
                  -g & 1   \end{array}\right]\left[\begin{array}{c}
                 \alpha \\
                  \beta   \end{array}\right]
           \end{aligned}
    \end{equation}
where the fact that only if $g > 0$ ($g\neq 0 $ for $G$ is a positive definite matrix) could $\dot{\mathbbm{V}}$ not be greater than 0 for arbitrary $\alpha$ and $\beta$ has been applied in derivation. Noticing that the property of $\mathcal{M}:=\left[\begin{array}{cc} g & -g \\ -g & 1   \end{array}\right]$ determines whether $\dot{\mathbbm{V}}$ is negative definite or semi-negative definite. Two cases would be discussed separately as follows:
\begin{enumerate}
    \item \emph{Case 1: $0<g<1$} \\
    The determinate of $\mathcal{M}$ is
    \begin{equation}
        \label{M}
        |M|= g-g^2 = g(1-g) >0
    \end{equation}
    which implies that $\mathcal{M}$ is positive definite, indicating that $\dot{\mathbbm{V}}$ is negative definite. So $\Phi^{[i]}$ for $i = 1,\ldots,n$ and $\Tilde{\Theta}$ could converge to zero asymptotically (see \emph{Theorem 4.10 in \cite{khalil2002nonlinear}}). Then the second requirement of \emph{Problem \ref{problem2}}  is satisfied. Besides, according to \emph{Theorem 1}, the first requirement is also meet, and \emph{Theorem 2} holds.
    \item \emph{Case 2: $g=1$} \\
    In this case, equation (\ref{V4dot1}) could be derived into
    \begin{align}\label{V4dot2}
             \dot{\mathbbm{V}}=&  -\sum_{i=1}^N G \Vert \nabla^P V^{[i]} \Vert^2 \nonumber \\
             &-\Vert \left[\begin{array}{ccc} \frac{\partial V^{[1]}}{\partial \theta^{[i]}} & \ldots&\frac{\partial V^{[n]}}{\partial \theta^{[n]}}\end{array} \right]^{\top} -k_c L \Tilde{\Theta} \Vert^2 \nonumber\\
            \leq & -\sum_{i=1}^N G \Vert \nabla^P V^{[i]} \Vert^2
 \end{align}
    So $\dot{\mathbbm{V}}$ is negative semi-definite. Based on LaSalle's invariance principle (Theorem 4.4 in \cite{khalil2002nonlinear}), the trajectories of the entire system will converge to the largest invariant set $\mathcal{\Lambda_+}$ in $\mathcal{B}:=\{\dot{\mathbbm{V}}=0\} \subseteq\{\Phi=0\}$. To obtain the largest invariant set $\mathcal{\Lambda_+}$ in $\mathcal{B}$, it is necessary to set $\dot{\Tilde{\Theta}}=\mathbf{0}$, which can be achieved by satisfying $D^{\top} L \tilde{\boldsymbol{\Theta}}=D^{\top} D D^{\top} \tilde{\boldsymbol{\Theta}}=0$. As a result, we obtain $D^{\top} \tilde{\boldsymbol{\Theta}}=0$. Moreover, with $\boldsymbol{\Phi}=0$, we have $\dot{\Theta}=\mathbf{0}$, ensuring that the largest invariant set $\mathcal{\Lambda_+}$ in $\mathcal{B}$ is
$
\mathcal{\Lambda_+}=\left\{ \Phi=0, D^{\top} \tilde{\boldsymbol{\Theta}}=\mathbf{0}\right\}.
$
Consequently, the errors in path following vanish asymptotically for all robots, and the differences of neighboring virtual coordinates $\tilde{\boldsymbol{\Theta}}$ converge to the desired formation pattern $\Delta^*$, thus achieving coordinated motion. It is noteworthy that (41) is positive definite and radially unbounded in $e$ and its time derivative is negative semi-definite, which guarantees the global vanishing of the composite error $\Tilde{\Theta}$ regardless of the initial composite error $\|\Tilde{\Theta}(t_0)\|$.
\end{enumerate}

\end{IEEEproof}
\begin{Remark}
    The cooperative requirement can be accomplished by selecting a convergence parameter $g$ that is less than or equal to one, but greater than zero. It is recommended that the value of $g$ is chosen appropriately in order to ensure a successful trajectory alignment with the intended path, while still satisfying the cooperative requirement. If the value of $g$ exceeds a certain threshold, it becomes apparent that the convergence rate is excessively rapid, rendering collaboration unfeasible.
\end{Remark}
\section{Applications for 2-D and 3-D cooperative moving path following problems}
\label{section5}
In the present section, we aim to showcase the efficacy of the previously discussed guiding vector field approach through its application to the problem of cooperative moving path following (CMPF). For a typical moving path following problem, the desired path is static with respect to a moving target. So there exists a contractual transformation between the original frame and the frame attached to the moving target. Contractual transformation is an isometric transformation that preserves the topological properties of a curve, and thus is a homotopy equivalence transformation. This demonstration starts from the theoretical analysis for 2-D situation, then attaching a simulation with an ellipse path situation, and ends with a 3-D cooperative moving path following simulation.

\subsection{Theoretical analysis for 2-D CMPF problem}
\subsubsection{Problem Formulation}
\begin{figure}[htb]
    \centering
    \includegraphics[width=1 \linewidth]{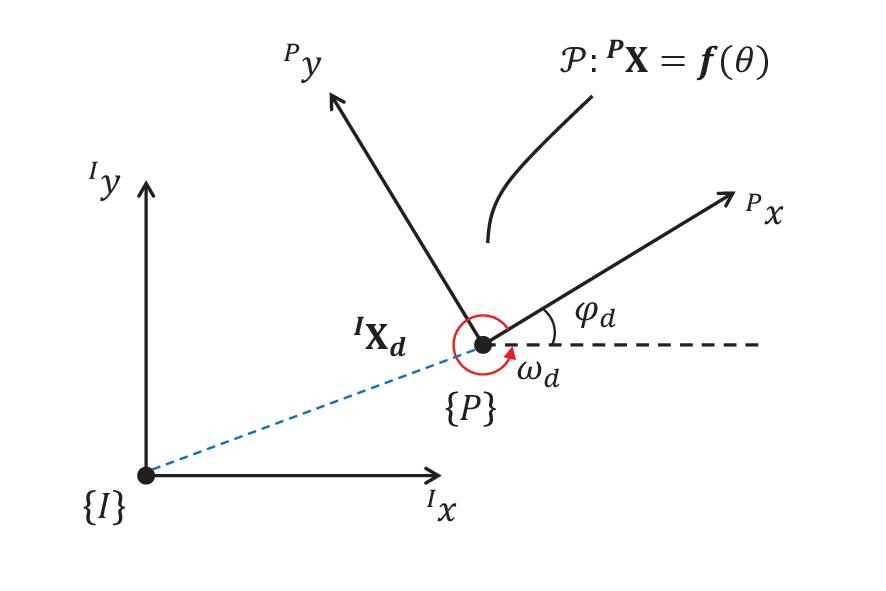}
    \caption{Relative position relationships among $\{I\}$ and $\{P\}$}
    \label{fig:5-1}
\end{figure}

Basically, there exists two frames, showed in figure \ref{fig:5-1}, in a moving path following problem. The former frame is the original frame ${I} = \{^Ix,^Iy\}$ attached to the earth and the latter ${P} = \{^Px,^Py\}$ attaching to a moving target with a kinetic model as
\begin{equation}
    \label{nh kinetic}
    \left\{\begin{aligned}
    &^I\dot{x}_d=v_d \cos \varphi_d, \\
    &^I\dot{y}_d=v_d \sin \varphi_d, \\
    & \dot{\varphi}_d=\omega_d
    \end{aligned}\right.
\end{equation}
where $(^Ix_d,^Iy_d,\,\varphi_d)$ is the position in frame $\{I\}$ and $(v_d,\omega_d)$ is the velocity and angular speed. The orientation of the axes are defined as follows: the $^Ix$ axis points to the East, the $^Iy$ points to the North, the $^Px$ axis points to the direction parallel with $v_d$ and $^Py$ is orthogonal with $^Px$ following a right hand rule. In practice, the moving target can be a leader robot or ship, and the desire path $^P\pmb{x}=\pmb{f}(\theta)$ is relative static with respect to the moving target. So if the moving path following problem was solved, the controlled agent would move along a fixed path in the frame attaching to the moving target.

The MPF problem can be expressed in the following manner: Consider a mobile robot that is in a motion described as $^I\dot{\pmb{x}}=\pmb{u}$ and a pre-defined path $\mathcal{P}:^P\pmb{x} = \pmb{f}(\theta)$. The objective is to devise a control strategy that guides the robot to follow the desired path $\mathcal{P}$ and maintains its trajectory on the same path.

\begin{Remark}
    The MPF problem has been addressed in previous studies, including those in \cite{Oliveira_2016} and \cite{Wang_2019} , which utilized the Frenet coordinate system based on the local structure of a curve. However, due to the global definition of the path, the Frenet description may give rise to singularities as the natural parameter moves along the path. To overcome this issue, an alternative approach is to use the guiding vector field method. In this research, we propose a non-singular guiding vector field solution to the MPF problem, which effectively circumvents the singularities that may arise from the Frenet framework.
\end{Remark}
\subsubsection{Guiding vector field approach for single agent}
\label{51b}
In a planar MPF problem, the transformation between $\{ I\}$ and $\{ P\}$ is
\begin{equation}
    \label{Ftrans}
    \mathcal{F} : \left[\begin{array}{c}^Px \\ ^Py \end{array}\right] = {}^{P}R_{I}(\theta_d)\left( \left[\begin{array}{c}^Ix \\ ^Iy \end{array}\right] - \left[\begin{array}{c}^Ix_d \\ ^Iy_d \end{array}\right] \right)
\end{equation}
The desired path $\mathcal{P}$ could be described as $^Px = f_1(\theta), ^Py = f_2(\theta)$ in frame $\{ P \}$. To stretch the 2-D curve into 3-D manifold, rewrite the equation as
\begin{equation}
    \begin{array}{cc}
         \phi_1 = {}^Px - f_1(\theta) \\
         \phi_2 = {}^Py - f_2(\theta)
    \end{array}
    \label{p-phi}
\end{equation}
Choose a positive definite candidate Lyapunov function as $V = k_1\frac{1}{2}\phi_1^2 + k_2\frac{1}{2}\phi_2^2$, where $k_1$ and $k_2$ are positive constants. The time derivative of $V$ is
\begin{equation}
    \label{Vdot1}
    \begin{aligned}
      \dot{V} = & \left[\begin{array}{cc} k_1\phi_1 &k_2\phi_2  \end{array} \right] \left[\begin{array}  {cc} \dot{\phi_1} \\ \dot{\phi_2}  \end{array} \right] \\
      =& \left[\begin{array}{cc} k_1\phi_1 &k_2\phi_2  \end{array} \right] \left[\begin{array}  {cc} \nabla ^{\xi}\phi_1^{\top} \\ \nabla ^{\xi}\phi_2^{\top}  \end{array} \right] \left[ \begin{array}  {ccc} ^P\dot{x} \\ ^P\dot{y}\\ \dot{\theta}   \end{array} \right] \\
      =& \left[\begin{array}{cc} k_1\phi_1 &k_2\phi_2  \end{array} \right] \left(\left[\begin{array}  {cc} \frac{\partial \phi_1}{\partial ^Px}& \frac{\partial \phi_1}{\partial ^Py}\\ \frac{\partial \phi_2}{\partial ^Px} & \frac{\partial \phi_2}{\partial ^Py} \end{array} \right] \left[ \begin{array}  {cc} ^P\dot{x} \\ ^P\dot{y}   \end{array} \right] + \left[ \begin{array}  {ccc} \frac{\partial \phi_1}{\partial \theta} \\ \frac{\partial \phi_2}{\partial \theta}   \end{array} \right] \dot{\theta} \right)
    \end{aligned}
\end{equation}
where $\nabla ^{\xi}\phi_i^{\top} =  \begin{bmatrix}\frac{\partial \phi_i}{\partial ^Px} & \frac{\partial \phi_i}{\partial ^Py} & \frac{\partial \phi_1}{\partial \theta}\end{bmatrix} $, noticing the fact that $ \left[\begin{array}  {cc} \frac{\partial \phi_1}{\partial ^Px}& \frac{\partial \phi_1}{\partial ^Py}\\ \frac{\partial \phi_2}{\partial ^Px} & \frac{\partial \phi_2}{\partial ^Py} \end{array} \right] = \left[\begin{array}  {cc} 1& 0\\ 0 & 1 \end{array} \right]$ concluded from (\ref{p-phi}), (\ref{Vdot1}) could be simplified as

\begin{equation}
    \label{Vdot2}
    \dot{V} = \left[\begin{array}{cc} k_1\phi_1 &k_2\phi_2  \end{array} \right] \left( \left[ \begin{array}  {cc} ^P\dot{x} \\ ^P\dot{y}   \end{array} \right] + \left[ \begin{array}  {ccc} \frac{\partial \phi_1}{\partial \theta} \\ \frac{\partial \phi_2}{\partial \theta}   \end{array} \right] \dot{\theta} \right)
\end{equation}

An inspiring fact about (\ref{Vdot2}) is that the extra parameter $\theta$ can be derived from the coordination $^Px $ and $ ^Py$. Thus the composition for time-varying transformation could be designed without any information about the third dimension.
\begin{myTheo}
    For robot with motion $^I\dot{\pmb{x}}=\pmb{u}$ and a pre-defined path $\mathcal{P}:^P\pmb{x} = \pmb{f}(\theta)$ with respect to a moving frame $\{P\}$, guiding vector field $\chi_{pf} = \left[ \begin{array}{cc}\pmb{u} & \dot{\theta}\end{array}\right]^T $ could accomplish the MPF problem with the following expression:
    \begin{equation}
        \label{5-2control input}
        \begin{aligned}
            &\left[ \begin{array}{c}\pmb{u} \\ \dot{\theta}\end{array}\right] =
            \left[ \begin{array}{c} ^I\dot{x}_d \\ ^I\dot{y}_d \\ 0 \end{array}\right]  -
            \left[ \begin{array}{cc} ^IR_P(\varphi_d)S(\omega_d) & \mathbf{0}_{2\times 1} \\ \mathbf{0}_{1\times 2} & 0  \end{array}\right] \left[ \begin{array}{c} ^Px \\ ^Py \\ 0 \end{array}\right] \\
            & + \left[ \begin{array}{cc} ^IR_P(\varphi_d) & \mathbf{0}_{2\times 1} \\ \mathbf{0}_{1\times 2} & 1  \end{array}\right] \left( \wedge(\nabla^{\xi}\phi_1,\nabla^{\xi}\phi_2)-\sum_{i=1}^2k_i\phi_i \nabla ^P\phi_i\right)
        \end{aligned}
    \end{equation}
    where $S(\omega_d) = \left[ \begin{array}{cc}
        0 & \omega_d \\
        -\omega_d & 0
    \end{array}\right]$ and the rotational matrix from $\{P\}$ to $\{I\}$ is $ ^IR_P(\varphi_d) = \left[ \begin{array}{cc}
        \cos{\varphi_d} & -\sin{\varphi_d} \\
        \sin{\varphi_d} & \cos{\varphi_d}
    \end{array}\right]$.
\end{myTheo}
\begin{IEEEproof}
    This proof starts from deriving the time derivative of $\left[ \begin{array}{cc}
        ^Ix & ^Iy
    \end{array}\right]^{\top}$. Concluding from (\ref{Ftrans}),
    \begin{align}
        \label{Ftrans_dot}
            \left[ \begin{array}{c}
        ^P\dot{x} \\ ^P\dot{y} \end{array}\right] &= S(\omega_d) \left[ \begin{array}{c}
             ^Px \\ ^Py
        \end{array}\right] \nonumber\\
        &\quad+ ^IR_P(\varphi_d)\left(\left[ \begin{array}{c} ^I\dot{x} \\ ^I\dot{y} \end{array}\right] - \left[ \begin{array}{c} ^I\dot{x}_d \\ ^I\dot{y}_d \end{array}\right] \right)
    \end{align}

    Denote the first two items in $\left( \wedge(\nabla^{\xi}\phi_1,\nabla^{\xi}\phi_2)-\Sigma_{i=1}^2k_i\phi_i \nabla ^{\xi}\phi_i\right)$ as $u_1$ and $u_2$, considering the fact $\left[ \begin{array}{c} ^I\dot{x}_d \\ ^I\dot{y}_d \end{array}\right] = \pmb{u}$, one obtains
        \begin{align}\label{Vdot3}
            \dot{V} =& \begin{bmatrix}k_1 \phi_1& k_2 \phi_2  \end{bmatrix} \left( \left[ \begin{array}  {cc} u_1 \\ u_2   \end{array} \right] + \left[ \begin{array}  {ccc} \frac{\partial \phi_1}{\partial \theta} \\ \frac{\partial \phi_2}{\partial \theta}   \end{array} \right] \dot{\theta} \right) \nonumber\\
            =& \begin{bmatrix}k_1 \phi_1& k_2 \phi_2  \end{bmatrix}\left[\begin{array}  {cc} \nabla ^{\xi}\phi_1^{\top}  \nonumber\\ \nabla ^{\xi}\phi_2^{\top}  \end{array} \right] (\wedge(\nabla^{\xi}\phi_1,\nabla^{\xi}\phi_2) \\ &-\sum_{i=1}^2k_i\phi_i \nabla ^{\xi}\phi_i )\nonumber\\
            =& -[k_1\phi_1 \nabla^{\xi}\phi_1^{\top} + k_2 \phi_2 \nabla^{\xi}\phi_2^{\top}]^2 \nonumber\\
            =& -\Vert \nabla^{\xi} V \Vert^2
        \end{align}
    meaning $\dot{V}$ is negative definite. All motions would go into the invariant set $\{ \phi_1 = 0, \phi_2 = 0\}$. In this invariant set, the coordination of the controlled agent satisfies $^Px = f_1(\theta)$ and $^Py = f_2(\theta)$. That is to say the position of robot finally converge to the path $\mathcal{P}$.
\end{IEEEproof}
\begin{Remark}
    The primary contribution of the initial two elements within the control law (\ref{5-2control input}) is to amalgamate the impact of the time-varying frame. This necessitates obtaining the position and velocity information of the moving target. In certain real-world scenarios, such as formation control with a leader-follower structure, this information can be evaluated from an internal model based on the data transmission between each distributed controller or observer.
\end{Remark}
\subsubsection{Cooperative guiding vector field for multi-agents}
    The preceding section \ref{51b} provides an overview of the guiding vector field that resolves the MPF challenge for a solitary agent. This segment, however, delves into a distributed formation controller, which is constructed based on the findings of reference \cite{yao2022guiding}. This controller serves as an effective tool to guide multiple agents in accomplishing the MPF task while optimizing their cooperation efforts.

    The main purpose for cooperative vector field is to design an extra mechanism with parameter $\theta^{[i]}$ to maintain a desired formation pattern $\Delta^{[i,j]}$ for $(i,j) \in \mathcal{E}$. Mathematically
    \begin{equation}
        \label{5-3pro}
        \lim_{t\rightarrow+\infty}(\theta^{[i]} - \theta^{[j]} - \Delta^{[i,j]})=0, (i,j)\in \mathcal{E}
    \end{equation}

    To solve the coordination problem (\ref{5-3pro}), the coordination vector field could be given as
    \begin{equation}
        \label{cr vector}
        \chi_{\mathrm{cr}}^{[i]} = \left( 0, 0 ,-\sum_{j\in\mathcal{N}_i} (\theta^{[i]} - \theta^{[j]} - \Delta^{[i,j]})  \right)^{\top}
    \end{equation}
    And the combined vector field could be given by $\mathfrak{X}^{[i]} = \chi_{\mathrm{mpf}}^{[i]}+\chi_{\mathrm{cr}}^{[i]}$. Noticing a fact that the last row of this vector field is the same as the vector field (4) given in \cite{yao2022guiding}, so the convergence is obvious.

\begin{Remark}
    From (\ref{5-2control input}) and (\ref{cr vector}), we can easily find a fact that the guiding part and the coordination part are separated, which decouples the design for vector field. This could be interpreted as the parameter $\theta_i$ goes outside the plane where the transformation occurs. That is also why we call the $\theta_i$ an extra dimension.
\end{Remark}

\subsection{A simulation for 2-D CMPF problem}
\begin{figure*} [t!]
	\centering
	\subfloat[\label{fig:3a}]{
		\includegraphics[width=0.5 \linewidth]{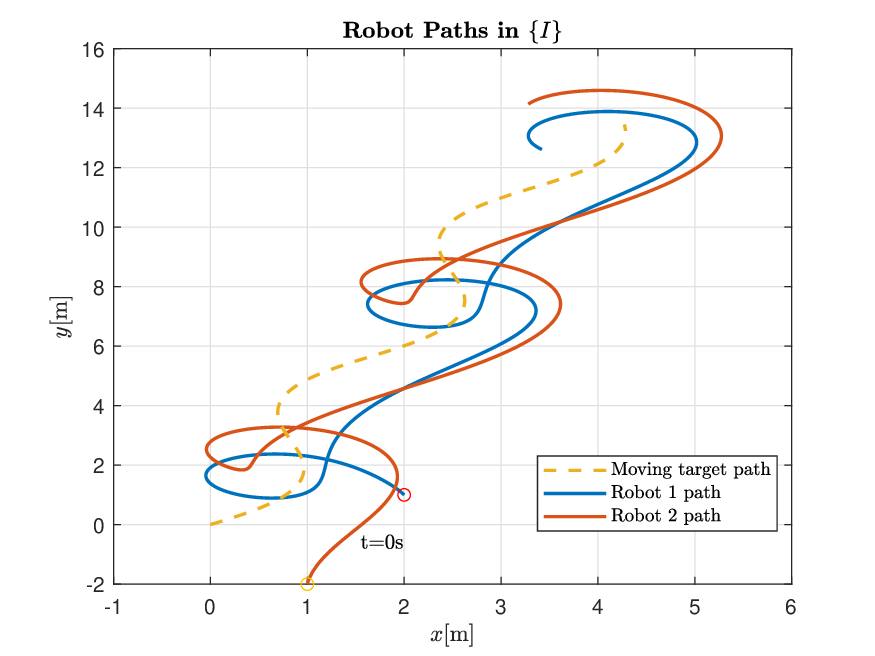}}
	\subfloat[\label{fig:3b}]{
		\includegraphics[width=0.5 \linewidth]{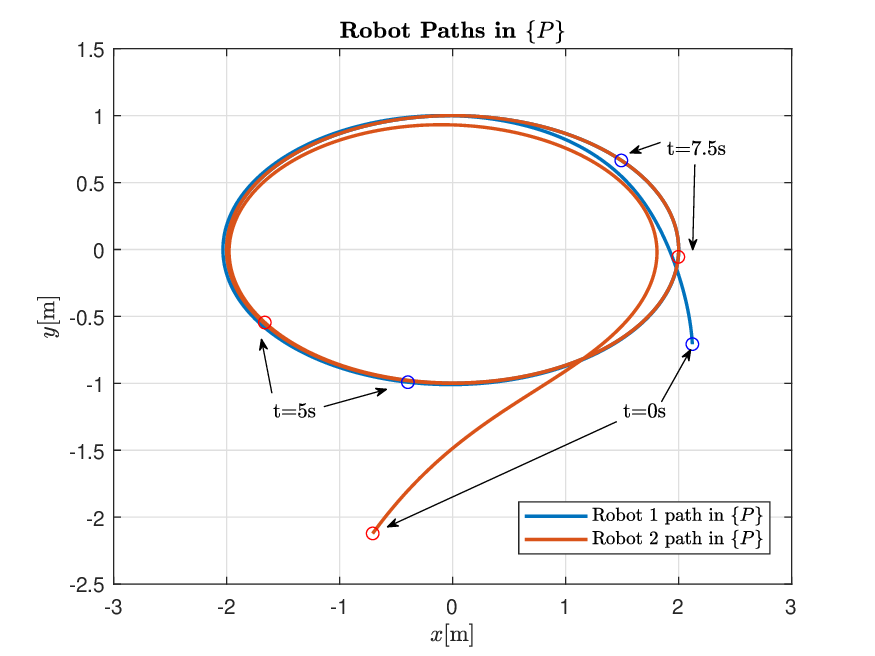}}
	\\
	\subfloat[\label{fig:3c}]{
		\includegraphics[width=0.5 \linewidth]{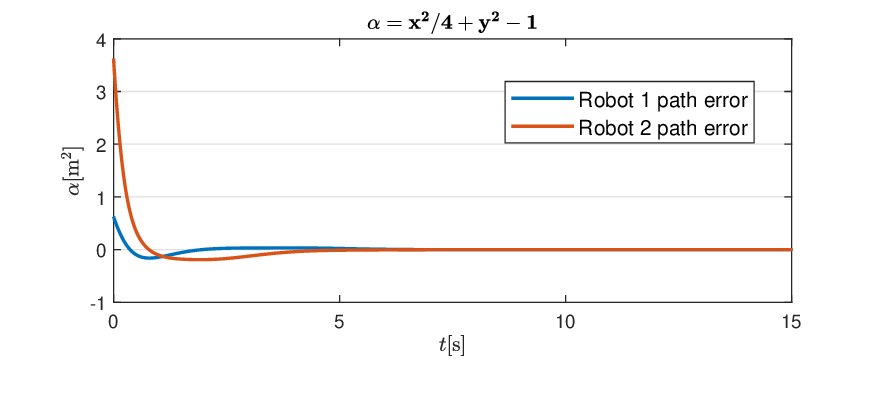} }
	\subfloat[\label{fig:3d}]{
		\includegraphics[width=0.5 \linewidth]{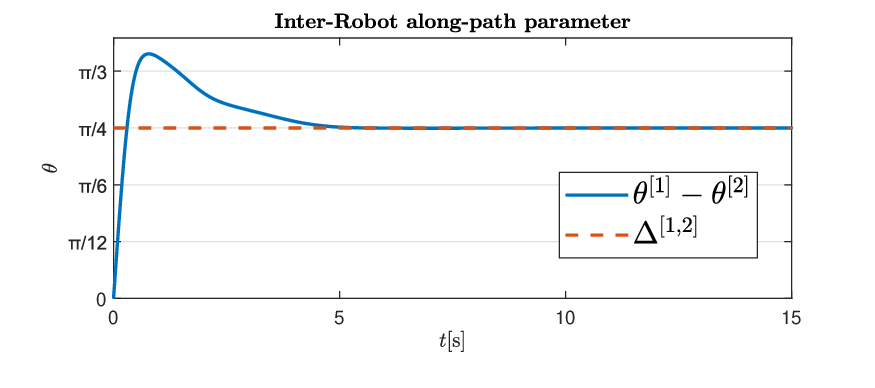}}
	\caption{Results of Simulation 1. (\ref{fig:3a}) shows the positions of the moving target and robots in $\{I\}$. (\ref{fig:3b}) shows the positions of the robots in $\{P\}$. (\ref{fig:3c}) shows the path errors of the robots. (\ref{fig:3d}) shows the coordination error between the robots.}
	\label{fig52}
\end{figure*}

In the first simulation, we let $N=2$ and the robots can get information from each other. The desired path has an ellipse shape $\alpha = x^2/4+y^2-1$, which can be parameterized as $x = 2\cos{\theta}, y = \sin{\theta}$. the motion of the moving target is
\begin{equation}
    \dot{x}_d = v_d\cos{\varphi}_d,\ \dot{y}_d = v_d\sin{\varphi}_d,\ \dot{\varphi}_d = \omega_d
\end{equation}
where $v_d = 1\mathrm{m/s}$ and $\omega_d = 0.5 \sin{t}\  \mathrm{rad/s^{-1}}$. The initial conditions for robot 1 and 2 are $(2\mathrm{m},1\mathrm{m})$ and $(1\mathrm{m},-2\mathrm{m})$, respectively. The formation pattern is set as $\Delta^{[1,2]}=\pi/4$. The vector field parameters are chosen as $k_1=k_2=k_c=1$. According to Figure.\ref{fig52}, all robots effectively track the ellipse shaped path while maintaining the desired positions relative to one another (as defined by $\theta^{[i]}$). Notably, both the path-following errors and coordination errors eventually dissipate towards zero.

\subsection{A simulation for 3-D CMPF problem}
The theoretical analysis of the 3-dimensional scenario parallels that of the 2-dimensional scenario, hence we shall present essential contextual data before showcasing the simulation results.

In this simulation, we assume the moving frame is attached on an aerial vehicle, whose kinetic model is described in \cite{Zuo_2022Unmanned}. Two frame in this case is the earth coordinate frame $\{I;{}^Ix,{}^Iy,{}^Iz\}$ and the aircraft-body coordinate frame $\{P;{}^Px,{}^Py,{}^Pz\}$. The position of the aircraft in $\{I\}$ is $^IX_d = (x_d,y_d,z_d)^{\top}$ and $(u,v,w)^{\top}$ is the inertial velocity of the aircraft measured in $\{P\}$. Here, Euler angles denoted as $(\psi_1,\psi_2,\psi_3)^{\top}$ following an $x-y-z$ rotation sequence and the Euler matrix is represented by $^PC_I = {}^PC_I(\psi_1,\psi_2,\psi_3)$. So the transformation between $\{I\}$ and $\{P\}$ is
\begin{equation}
    \label{3-D transformation}
    ^PX = {}^PC_I({}^IX-{}^IX_d)
\end{equation}
where $^IX_d$ has the following transnational kinematics
\begin{equation}
    \label{transnational kinematics of airship}
    \left\{
    \begin{aligned}
    \dot{x}_d= & u \cos \psi_2 \cos \psi_3+v(\sin \psi_1 \sin \psi_2 \cos \psi_3-\cos \phi \sin \psi_3) \\
    & +w(\sin \psi_1 \sin \psi_3+\cos \phi \sin \psi_2 \cos \psi_3) \\
    \dot{y}_d= & u \cos \psi_2 \sin \psi_3+v(\sin \psi_1 \sin \psi_2 \sin \psi_3+\cos \phi \cos \psi_3) \\
    & -w(\sin \psi_1 \cos \psi_3-\cos \phi \sin \psi_2 \sin \psi_3) \\
    \dot{z}_d= & -u \sin \psi_2+v \sin \psi_1 \cos \psi_2+w \cos \phi \cos \psi_2
    \end{aligned}\right.
\end{equation}
Besides, the rotational equations of the aircraft in relation to Euler angles are denoted as follows:
\begin{equation}
    \label{rotational equations of airship}
    \left\{\begin{array}{l}
        \dot{\psi}_1=p+(r \cos \psi_1+q \sin \psi_1) \tan \psi_2 \\
        \dot{\psi}_2=q \cos \psi_1-r \sin \psi_1 \\
        \dot{\psi}_3=\frac{1}{\cos \psi_2}(r \cos \psi_1+q \sin \psi_1)
    \end{array}\right.
\end{equation}
where $[p,q,r]^{\top}$ is the angular velocity vector of the aircraft measured in the frame $\{P\}$.
\begin{Remark}
    In rotational kinematics, the Euler angles representation commonly encounters singular points. This phenomenon is attributed to the inherent nature of the description, rather than the guiding vector field approach. Therefore, in the context of this simulation, it is imperative to assign negligible values to the angular velocities.
\end{Remark}

\begin{figure}
    \centering
    \includegraphics{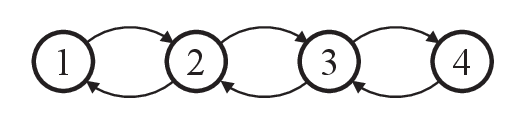}
    \caption{The topology in simulation 2}
    \label{fig:f54}
\end{figure}

In this simulation, we let $N=4$ and the communication topology  is shown in Fig. \ref{fig:f54}. The desired path here is a Lissajous curve, expressing as $^Px = 2\cos{\theta}$, $^Py = \sin{\theta}$ and $^Pz = \cos{\frac{\theta}{2}}$. The speed and angular velocity for the aircraft is set as $(u,v,w)^{\top} = (1+0.1\sin{t} \  \mathrm{m/s},0.1\cos{t} \  \mathrm{m/s},0.1\sin{t}  \ \mathrm{m/s})^{\top}$ and $(p,q,r)^{\top} = (\frac{0.01\pi}{180} \sin{t} \ \mathrm{rad/s},\frac{0.01\pi}{180} \sin{t} \  \mathrm{rad/s},\frac{0.01\pi}{180} \sin{t} \ \mathrm{rad/s})^{\top}$ with initial conditions set as $(x_{d0},y_{d0},z_{d0})^{\top} = (0,0,1  \mathrm{m})^{\top}$ and $(\psi_{10},\psi_{20},\psi_{30})^{\top} = (0,0,\pi/4)^{\top}$, respectively. The initial conditions for each aerial vehicle are $(1  \mathrm{m},0,0)^{\top}$, $(1  \mathrm{m},1  \mathrm{m},0)^{\top}$, $(-1  \mathrm{m},0,0)^{\top}$ and $(1  \mathrm{m},0,2  \mathrm{m})^{\top}$. The formation pattern is set as $\Delta^{[1,2]} =\Delta^{[2,3]} =\Delta^{[3,4]} = \pi/6$. Besides, the vector field parameters are set as $k_1 = k_2 = k_3 = 1$ and $k_c = 5$. $ \Vert \Phi \Vert^2 = ({}^Px-2\cos{\theta})^2 + ({}^Py-\sin{\theta})^2+({}^Pz-\cos{\frac{\theta}{2}})^2 $ is used to described the convergence error.

\begin{figure*} [t!]
	\centering
	\subfloat[\label{fig:5a}]{
		\includegraphics[width=0.5 \linewidth]{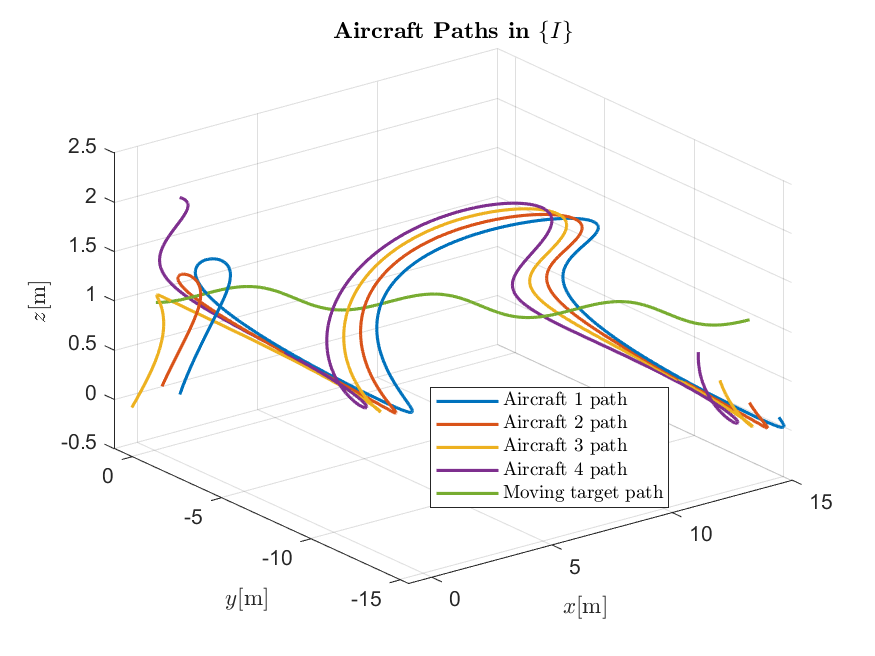}}
	\subfloat[\label{fig:5b}]{
		\includegraphics[width=0.5 \linewidth]{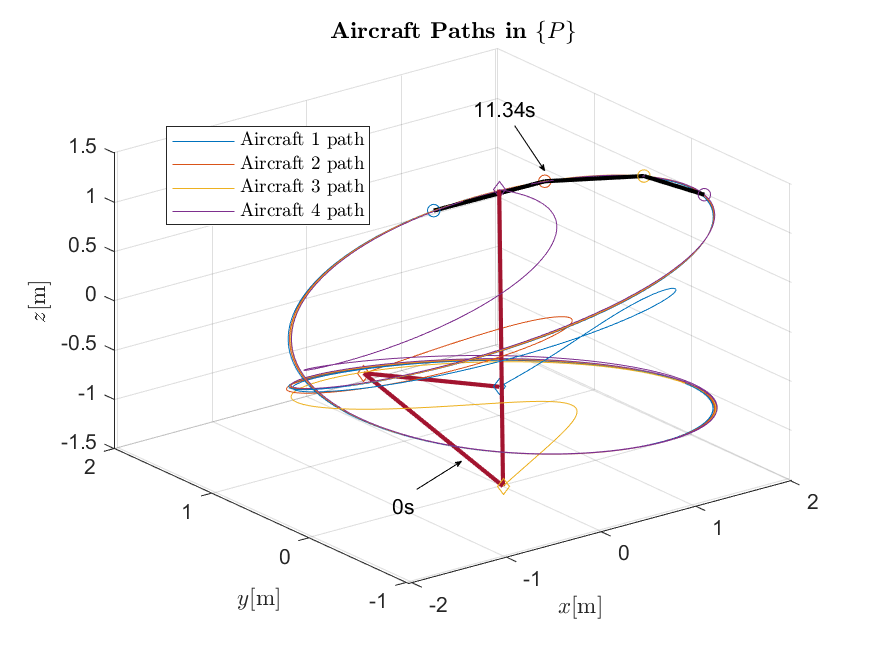}}
	\\
	\subfloat[\label{fig:5c}]{
		\includegraphics[width=0.5 \linewidth]{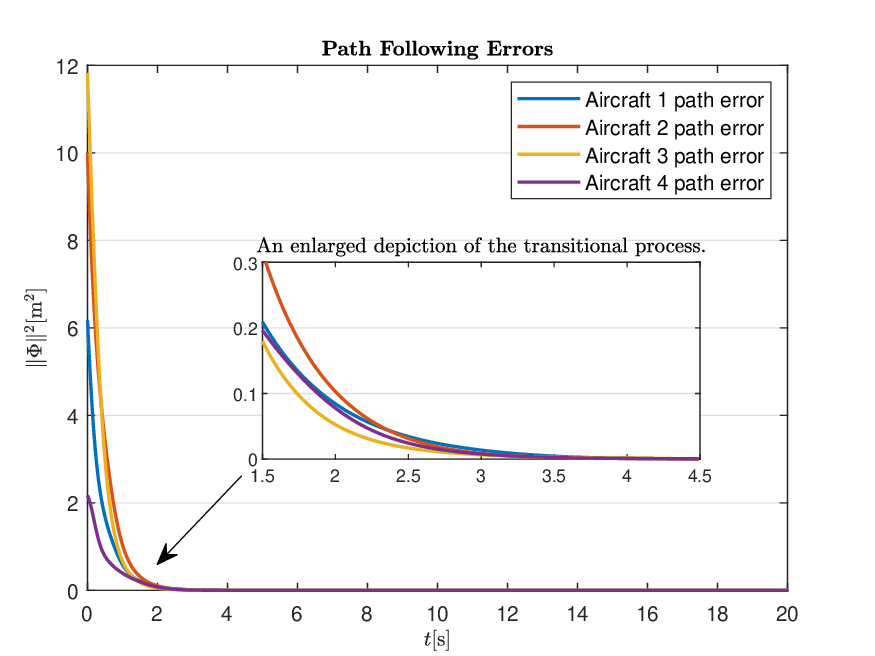} }
	\subfloat[\label{fig:5d}]{
		\includegraphics[width=0.5 \linewidth]{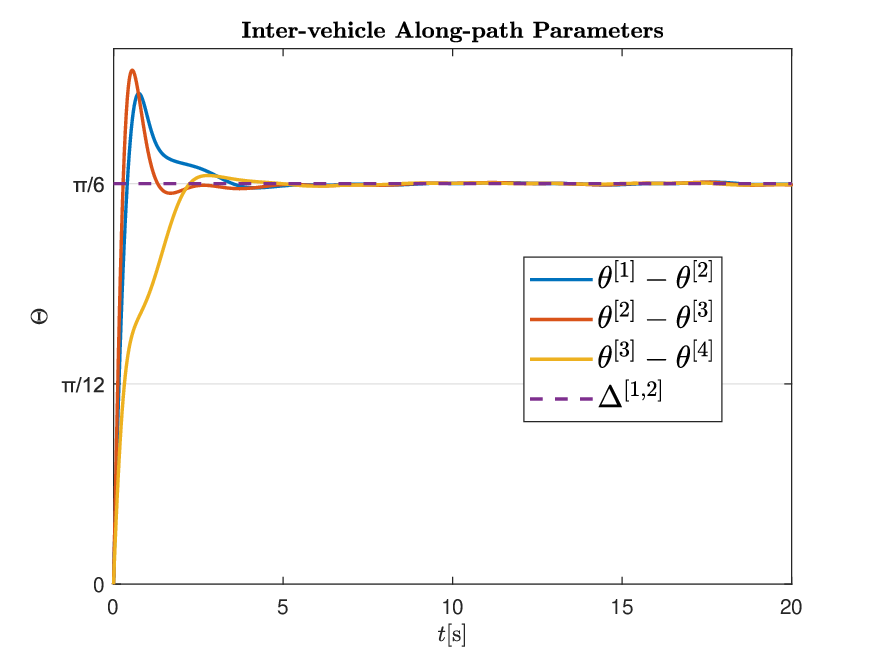}}
	\caption{Results of Simulation 2. (\ref{fig:5a}) shows the positions of the moving target and robots in $\{I\}$. (\ref{fig:5b}) shows the positions of the robots in $\{P\}$m, and the formation at $t=11.34\mathrm{s}$ has been marked. (\ref{fig:5c}) shows the path errors of the robots. (\ref{fig:5d}) shows the coordination error between the robots.}
	\label{fig54}
\end{figure*}

Fig. \ref{fig54} shows the simulation results. In due course, both the errors related to path-following and coordination are inclined to reduce and ultimately converge towards zero.

\section{Conclusion and Future Work}
\label{section 6}
This article introduces the concept of a non-singular cooperative guiding vector field, achieved via a homotopy equivalence transformation. It firstly provides an elaboration on the derivation of a guiding vector field, based on a non-singular vector field, for navigating a time-varying transformed path from another frame. The article also deliberates on the existence of such vector fields, finding a necessary condition for such vector field is that the Jocabian of the transformation should be revertible. Subsequently, it presents a coordination vector field derived from the guiding vector field. Additionally, we discuss about the influence from the vector field parameters, finding that the convergence parameter should be limited ($0<g \leq1$) when taking the coordination vector field into consideration. Finally, the practical implementation of this innovative vector field is demonstrated through its application to a planar cooperative moving path following challenge, establishing its effectiveness.

Based on our research, we have identified several potential avenues for future work. Firstly, while the vector field in cooperative control typically exhibits identical properties, there has been limited investigation into the potential of non-identical vector fields to facilitate coordination. Secondly, with the increasing interest in heterogeneous multi-agent systems, exploring the feasibility of utilizing guiding vector field approaches in this area holds particular promise. Lastly, our research has demonstrated that guiding vector fields can function under homotopy equivalence transformation. However, further exploration is required to fully understand the necessary conditions for this approach.

% or


\appendices
\section{Proof of Lemma 2 }
To prove that the direct sum $\mathcal{F}^{ex} = \mathcal{F}\oplus \mathcal{A}$ of a homotopy equivalence transformation $\mathcal{F}$ and an identical transformation $\mathcal{A}$ is also a homotopy equivalence transformation, we need to show that $\mathcal{F}^{ex}$ satisfies the following two conditions:

$\mathcal{F}^{ex}$ is a continuous transformation.

$\mathcal{F}^{ex}$ and its inverse are homotopy equivalences.

To prove the first condition, note that the direct sum $\mathcal{F}^{ex}$ is defined as the transformation that acts on the direct sum space of the spaces operated by $\mathcal{F}$ and $\mathcal{A}$. Since both $\mathcal{F}$ and $\mathcal{A}$ are continuous transformations, it follows that $\mathcal{F}^{ex}$ is also continuous.

To prove the second condition, we need to show that $\mathcal{F}^{ex}$ and its inverse are homotopy equivalences. Let $\mathcal{G}$ be the inverse of $\mathcal{F}$, i.e., $\mathcal{F}\circ \mathcal{G} = \mathcal{A}$ and $\mathcal{G}\circ \mathcal{F} = \mathcal{A}$. Then, the direct sum of $\mathcal{G}$ and $\mathcal{A}$, denoted as $\mathcal{G}^{ex} = \mathcal{G}\oplus \mathcal{A}$, is the inverse of $\mathcal{F}^{ex}$, i.e., $\mathcal{F}^{ex}\circ \mathcal{G}^{ex} = \mathcal{A}$ and $\mathcal{G}^{ex}\circ \mathcal{F}^{ex} = \mathcal{A}$.

Next, we need to show that $\mathcal{F}^{ex}$ and $\mathcal{G}^{ex}$ are homotopy equivalences. To do so, we will construct a homotopy between $\mathcal{F}^{ex}\circ \mathcal{G}^{ex}$ and $\mathcal{A}$, and a homotopy between $\mathcal{G}^{ex}\circ \mathcal{F}^{ex}$ and $\mathcal{A}$.

For the first homotopy, consider the homotopy $H_1: [0,1]\times (\mathcal{F}\times \mathcal{G})\times \mathcal{A} \rightarrow \mathcal{F}\oplus \mathcal{A}$ defined by $H_1(t,(f,g),a) = (f(ta), a)$, where $f\in \mathcal{F}$, $g\in \mathcal{G}$, $a\in \mathcal{A}$, and $t\in [0,1]$. Note that $H_1(0,(f,g),a) = (f(0),a) = (a,a) = H_1(1,(f,g),a)$ for all $(f,g)\in (\mathcal{F}\times \mathcal{G})$ and $a\in \mathcal{A}$, and $H_1(t,(f,g),a) \in \mathcal{F}\oplus \mathcal{A}$ for all $t\in [0,1]$ and $(f,g,a)\in (\mathcal{F}\times \mathcal{G})\times \mathcal{A}$. Therefore, $H_1$ is a well-defined homotopy between $\mathcal{F}^{ex}\circ \mathcal{G}^{ex}$ and $\mathcal{A}$.

For the second homotopy, consider the homotopy $H_2: [0,1]\times (\mathcal{G}\times \mathcal{F})\times \mathcal{A} \rightarrow \mathcal{G}\oplus \mathcal{A}$ defined by $H_2(t,(g,f),a) = (g(ta), a)$, where $g\in \mathcal{G}$, $f\in \mathcal{F}$, $a\in \mathcal{A}$, and $t\in [0,1]$. Note that $H_2(0,(g,f),a) = (g(0),a) = (a,a) = H_2(1,(g,f),a)$ for all $(g,f)\in (\mathcal{G}\times \mathcal{F})$ and $a\in \mathcal{A}$, and $H_2(t,(g,f),a) \in \mathcal{G}\oplus \mathcal{A}$ for all $t\in [0,1]$ and $(g,f,a)\in (\mathcal{G}\times \mathcal{F})\times \mathcal{A}$. Therefore, $H_2$ is a well-defined homotopy between $\mathcal{G}^{ex}\circ \mathcal{F}^{ex}$ and $\mathcal{A}$.

Since we have constructed homotopies between $\mathcal{F}^{ex}\circ \mathcal{G}^{ex}$ and $\mathcal{A}$, and between $\mathcal{G}^{ex}\circ \mathcal{F}^{ex}$ and $\mathcal{A}$, we conclude that $\mathcal{F}^{ex}$ and $\mathcal{G}^{ex}$ are homotopy equivalences. Therefore, the direct sum $\mathcal{F}^{ex} = \mathcal{F}\oplus \mathcal{A}$ is also a homotopy equivalence transformation.

%\section{Proof of Theorem }

%\section*{Acknowledgment}


\ifCLASSOPTIONcaptionsoff
  \newpage
\fi



%\bibliographystyle{IEEEtran}
\bibliographystyle{IEEEtran}
\bibliography{references}{}

\begin{thebibliography}{10}
\providecommand{\url}[1]{#1}
\csname url@samestyle\endcsname
\providecommand{\newblock}{\relax}
\providecommand{\bibinfo}[2]{#2}
\providecommand{\BIBentrySTDinterwordspacing}{\spaceskip=0pt\relax}
\providecommand{\BIBentryALTinterwordstretchfactor}{4}
\providecommand{\BIBentryALTinterwordspacing}{\spaceskip=\fontdimen2\font plus
\BIBentryALTinterwordstretchfactor\fontdimen3\font minus
  \fontdimen4\font\relax}
\providecommand{\BIBforeignlanguage}[2]{{%
\expandafter\ifx\csname l@#1\endcsname\relax
\typeout{** WARNING: IEEEtran.bst: No hyphenation pattern has been}%
\typeout{** loaded for the language `#1'. Using the pattern for}%
\typeout{** the default language instead.}%
\else
\language=\csname l@#1\endcsname
\fi
#2}}
\providecommand{\BIBdecl}{\relax}
\BIBdecl

\bibitem{Oliveira_2016}
T.~Oliveira, A.~P. Aguiar, and P.~Encarnacao, ``\BIBforeignlanguage{en}{Moving
  path following for unmanned aerial vehicles with applications to single and
  multiple target tracking problems},'' \emph{\BIBforeignlanguage{en}{IEEE
  Transactions on Robotics}}, vol.~32, no.~5, pp. 1062--1078, 2016.

\bibitem{Wang_2019}
Y.~Wang, D.~Wang, and S.~Zhu, ``\BIBforeignlanguage{en}{Cooperative moving path
  following for multiple fixed-wing unmanned aerial vehicles with speed
  constraints},'' \emph{\BIBforeignlanguage{en}{Automatica}}, vol. 100, pp.
  82--89, 2019.

\bibitem{Sujit_2014}
P.~Sujit, S.~Saripalli, and J.~B. Sousa, ``\BIBforeignlanguage{en}{Unmanned
  aerial vehicle path following: A survey and analysis of algorithms for
  fixed-wing unmanned aerial vehicless},'' \emph{\BIBforeignlanguage{en}{IEEE
  Control Systems}}, vol.~34, no.~1, pp. 42--59, 2014.

\bibitem{Ostertag_2008An}
E.~Ostertag, ``\BIBforeignlanguage{en}{An improved path-following method for
  mixed ${H} _{2} /{H}_{\infty}$ controller design},''
  \emph{\BIBforeignlanguage{en}{IEEE Transactions on Automatic Control}},
  vol.~53, no.~8, pp. 1967--1971, 2008.

\bibitem{Caharija_2015A}
W.~Caharija, K.~Y. Pettersen, P.~Calado, and J.~Braga,
  ``\BIBforeignlanguage{en}{A comparison between the ilos guidance and the
  vector field guidance},'' \emph{\BIBforeignlanguage{en}{IFAC-PapersOnLine}},
  vol.~48, no.~16, pp. 89--94, 2015.

\bibitem{Borhaug_2008Integral}
E.~Borhaug, A.~Pavlov, and K.~Y. Pettersen, ``\BIBforeignlanguage{en}{Integral
  los control for path following of underactuated marine surface vessels in the
  presence of constant ocean currents},'' in \emph{\BIBforeignlanguage{en}{2008
  47th IEEE Conference on Decision and Control}}.\hskip 1em plus 0.5em minus
  0.4em\relax IEEE, 2008.

\bibitem{Caharija_2016Integral}
W.~Caharija, K.~Y. Pettersen, M.~Bibuli, P.~Calado, E.~Zereik, J.~Braga, J.~T.
  Gravdahl, A.~J. Sorensen, M.~Milovanovic, and G.~Bruzzone,
  ``\BIBforeignlanguage{en}{Integral line-of-sight guidance and control of
  underactuated marine vehicles: Theory, simulations, and experiments},''
  \emph{\BIBforeignlanguage{en}{IEEE Transactions on Control Systems
  Technology}}, vol.~24, no.~5, pp. 1623--1642, 2016.

\bibitem{Lapierre_2007Nonlinear}
L.~Lapierre and D.~Soetanto, ``\BIBforeignlanguage{en}{Nonlinear path-following
  control of an auv},'' \emph{\BIBforeignlanguage{en}{Ocean Engineering}},
  vol.~34, no. 11--12, pp. 1734--1744, 2007.

\bibitem{Faulwasser_2016Nonlinear}
T.~Faulwasser and R.~Findeisen, ``\BIBforeignlanguage{en}{Nonlinear model
  predictive control for constrained output path following},''
  \emph{\BIBforeignlanguage{en}{IEEE Transactions on Automatic Control}},
  vol.~61, no.~4, pp. 1026--1039, 2016.

\bibitem{Nakai_2013Vector}
K.~Nakai and K.~Uchiyama, ``\BIBforeignlanguage{en}{Vector fields for uav
  guidance using potential function method for formation flight},'' in
  \emph{\BIBforeignlanguage{en}{AIAA Guidance, Navigation, and Control (GNC)
  Conference}}.\hskip 1em plus 0.5em minus 0.4em\relax Reston, Virginia:
  American Institute of Aeronautics and Astronautics, 2013.

\bibitem{Rezende_2022}
A.~M.~C. Rezende, V.~M. Goncalves, and L.~C.~A. Pimenta,
  ``\BIBforeignlanguage{en}{Constructive time-varying vector fields for robot
  navigation},'' \emph{\BIBforeignlanguage{en}{IEEE Transactions on Robotics}},
  vol.~38, no.~2, pp. 852--867, 2022.

\bibitem{Lan_2011Synthesis}
Y.~Lan, G.~Yan, and Z.~Lin, ``\BIBforeignlanguage{en}{Synthesis of distributed
  control of coordinated path following based on hybrid approach},''
  \emph{\BIBforeignlanguage{en}{IEEE Transactions on Automatic Control}},
  vol.~56, no.~5, pp. 1170--1175, 2011.

\bibitem{Xie_2022Cooperative}
W.~Xie, D.~Cabecinhas, R.~Cunha, and C.~Silvestre,
  ``\BIBforeignlanguage{en}{Cooperative path following control of multiple
  quadcopters with unknown external disturbances},''
  \emph{\BIBforeignlanguage{en}{IEEE Transactions on Systems, Man, and
  Cybernetics: Systems}}, vol.~52, no.~1, pp. 667--679, 2022.

\bibitem{Alessandretti_2020An}
A.~Alessandretti and A.~P. Aguiar, ``\BIBforeignlanguage{en}{An
  optimization-based cooperative path-following framework for multiple robotic
  vehicles},'' \emph{\BIBforeignlanguage{en}{IEEE Transactions on Control of
  Network Systems}}, vol.~7, no.~2, pp. 1002--1014, 2020.

\bibitem{Eek_2021Formation}
{\AA}.~Eek, K.~Y. Pettersen, E.-L.~M. Ruud, and T.~R. Krogstad,
  ``\BIBforeignlanguage{en}{Formation path following control of underactuated
  usvs},'' \emph{\BIBforeignlanguage{en}{European Journal of Control}},
  vol.~62, pp. 171--184, 2021.

\bibitem{Yao_2021Singularity}
W.~Yao, H.~G. de~Marina, B.~Lin, and M.~Cao,
  ``\BIBforeignlanguage{en}{Singularity-free guiding vector field for robot
  navigation},'' \emph{\BIBforeignlanguage{en}{IEEE Transactions on Robotics}},
  pp. 1--16, 2021.

\bibitem{Nelson_2007}
D.~Nelson, D.~Barber, T.~McLain, and R.~Beard, ``\BIBforeignlanguage{en}{Vector
  field path following for miniature air vehicles},''
  \emph{\BIBforeignlanguage{en}{IEEE Transactions on Robotics}}, vol.~23,
  no.~3, pp. 519--529, 2007.

\bibitem{Goncalves_2010Vector}
V.~M. Goncalves, L.~C.~A. Pimenta, C.~A. Maia, B.~C.~O. Dutra, and G.~A.~S.
  Pereira, ``\BIBforeignlanguage{en}{Vector fields for robot navigation along
  time-varying curves in $n$-dimensions},'' \emph{\BIBforeignlanguage{en}{IEEE
  Transactions on Robotics}}, vol.~26, no.~4, pp. 647--659, 2010.

\bibitem{Liang_2014}
Y.~Liang, Y.~Jia, J.~Du, and F.~Matsuno, ``\BIBforeignlanguage{en}{Cooperative
  bicircular target tracking using multiple unmanned aerial vehicles},'' in
  \emph{\BIBforeignlanguage{en}{53rd IEEE Conference on Decision and
  Control}}.\hskip 1em plus 0.5em minus 0.4em\relax IEEE, 2014.

\bibitem{Kapitanyuk_2017}
Y.~A. Kapitanyuk, H.~G.~d. Marina, A.~V. Proskurnikov, and M.~Cao,
  ``\BIBforeignlanguage{en}{Guiding vector field algorithm for a moving path
  following problem},'' \emph{\BIBforeignlanguage{en}{IFAC-PapersOnLine}},
  vol.~50, no.~1, pp. 6983--6988, 2017.

\bibitem{Yao_2023}
W.~Yao, B.~Lin, B.~D.~O. Anderson, and M.~Cao,
  ``\BIBforeignlanguage{en}{Topological analysis of vector-field guided path
  following on manifolds},'' \emph{\BIBforeignlanguage{en}{IEEE Transactions on
  Automatic Control}}, vol.~68, no.~3, pp. 1353--1368, 2023.

\bibitem{Kapitanyuk_2018}
Y.~A. Kapitanyuk, A.~V. Proskurnikov, and M.~Cao, ``\BIBforeignlanguage{en}{A
  guiding vector-field algorithm for path-following control of nonholonomic
  mobile robots},'' \emph{\BIBforeignlanguage{en}{IEEE Transactions on Control
  Systems Technology}}, vol.~26, no.~4, pp. 1372--1385, 2018.

\bibitem{yao2022guiding}
W.~Yao, H.~G. de~Marina, Z.~Sun, and M.~Cao, ``Guiding vector fields for the
  distributed motion coordination of mobile robots,'' \emph{IEEE Transactions
  on Robotics}, 2022.

\bibitem{Zuo_2015Three}
Z.~Zuo, V.~Cichella, M.~Xu, and N.~Hovakimyan,
  ``\BIBforeignlanguage{en}{Three-dimensional coordinated path-following
  control for second-order multi-agent networks},''
  \emph{\BIBforeignlanguage{en}{Journal of the Franklin Institute}}, vol. 352,
  no.~9, pp. 3858--3872, 2015.

\bibitem{Zuo_2022}
Z.~Zuo, J.~Song, and Q.-L. Han, ``\BIBforeignlanguage{en}{Coordinated planar
  path-following control for multiple nonholonomic wheeled mobile robots},''
  \emph{\BIBforeignlanguage{en}{IEEE Transactions on Cybernetics}}, vol.~52,
  no.~9, pp. 9404--9413, 2022.

\bibitem{Frew_2008Coordinated}
E.~W. Frew, D.~A. Lawrence, and S.~Morris,
  ``\BIBforeignlanguage{en}{Coordinated standoff tracking of moving targets
  using lyapunov guidance vector fields},''
  \emph{\BIBforeignlanguage{en}{Journal of Guidance, Control, and Dynamics}},
  vol.~31, no.~2, pp. 290--306, 2008.

\bibitem{Olfati-Saber_2007}
R.~Olfati-Saber, J.~A. Fax, and R.~M. Murray,
  ``\BIBforeignlanguage{en}{Consensus and cooperation in networked multi-agent
  systems},'' \emph{\BIBforeignlanguage{en}{Proceedings of the IEEE}}, vol.~95,
  no.~1, pp. 215--233, 2007.

\bibitem{khalil2002nonlinear}
H.~Khalil, \emph{Nonlinear Systems}, ser. Pearson Education.\hskip 1em plus
  0.5em minus 0.4em\relax Prentice Hall, 2002.

\bibitem{Zuo_2022Unmanned}
Z.~Zuo, C.~Liu, Q.-L. Han, and J.~Song, ``\BIBforeignlanguage{en}{Unmanned
  aerial vehicles: Control methods and future challenges},''
  \emph{\BIBforeignlanguage{en}{IEEE/CAA Journal of Automatica Sinica}},
  vol.~9, no.~4, pp. 601--614, 2022.

\end{thebibliography}

%\end{IEEEbiography}
\begin{IEEEbiography}[{\includegraphics[width=1in,height=1.25in,clip,keepaspectratio]{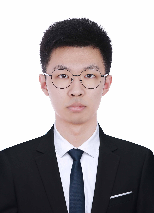}}]{Zirui Chen} received the B.Eng. degree in automation from Beihang University (BUAA), Beijing, China, in 2021, where he is currently pursuing the M.A.Eng. degree in control theory and applications. 

His research interests are in the fields of nonlinear system control and geometric control.
\end{IEEEbiography}

\begin{IEEEbiography}[{\includegraphics[width=1in,height=1.25in,clip,keepaspectratio]{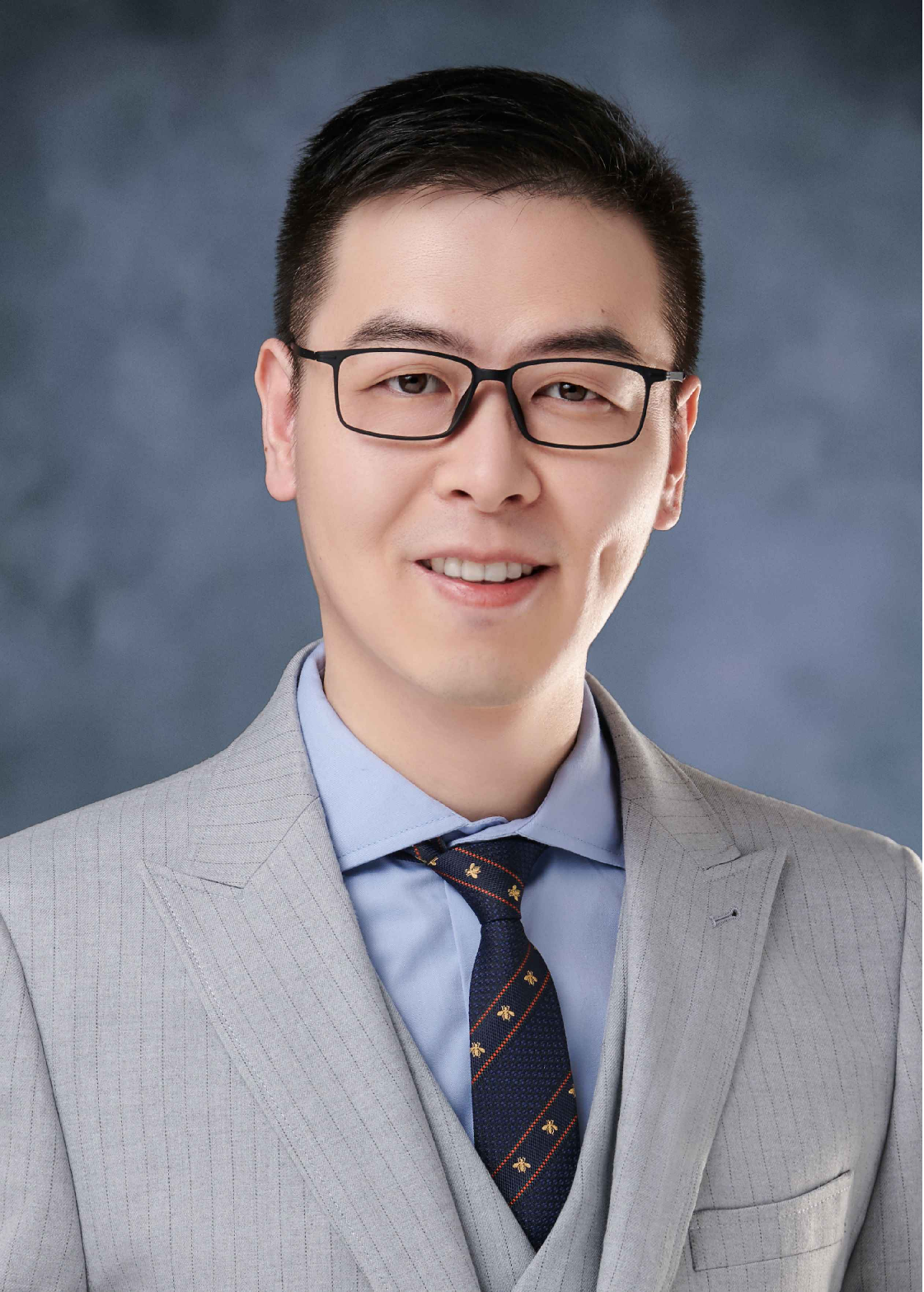}}]{Zongyu Zuo} (Senior Member, IEEE) received his B.Eng. degree in Automatic Control from Central South University, Hunan, China, in 2005, and Ph.D. degree in Control Theory and Applications from Beihang University (BUAA), Beijing, China, in 2011.

He was an academic visitor at the School of Electrical and Electronic Engineering, University of Manchester from September 2014 to September 2015 and held an inviting associate professorship at Mechanical Engineering and Computer Science, UMR CNRS 8201, Universit\'{e} de Valenciennes et du Hainaut-Cambr\'{e}sis in October 2015 and May 2017. He is currently a full professor at the School of Automation Science and Electrical Engineering, Beihang University. His research interests are in the fields of nonlinear system control, control of UAVs, and coordination of multi-agent system. He was identified as a Highly Cited Researcher - 2020 and 2022 by Clarivate Analytics as well as a most cited Chinese Researcher - 2021 and 2022  by Elsevier.

Prof. Zuo currently serves as an Associate Editor for IEEE Transactions on Industrial Informatics, IEEE/CAA Journal of Automatica Sinica, Journal of The Franklin Institute, Journal of Vibration and Control, and International Journal of Aeronautical {\&} Space Sciences.
\end{IEEEbiography}

%\end{IEEEbiography}


%\newpage



%\vfill


\end{document}